%% file: ParamEstim.tex
\documentclass[prd,aps,showpacs,superscriptaddress,nofootinbib,eqsecnum,amsfonts,amsmath, 
preprintnumbers,twocolumn,floatfix,floats]{revtex4}

\usepackage{epsfig}
\usepackage{bm}
\usepackage{graphics}
\usepackage{graphicx}
\usepackage{bm}
\usepackage{amssymb}
\usepackage[usenames]{color}

\def\bea{\begin{eqnarray}}
\def\eea{\end{eqnarray}}
\def\vt{\vartheta}

\begin{document}

\newcommand{\rhat}{\hat{r}}
\newcommand{\iotahat}{\hat{\iota}}
\newcommand{\phihat}{\hat{\phi}}
\newcommand{\h}{\mathfrak{h}}
\newcommand{\be}{\begin{equation}}
\newcommand{\ee}{\end{equation}}
\newcommand{\ber}{\begin{eqnarray}}
\newcommand{\eer}{\end{eqnarray}}
\newcommand{\fmerg}{f_{\rm merg}}
\newcommand{\fcut}{f_{\rm cut}}
\newcommand{\fring}{f_{\rm ring}}
\newcommand{\cA}{\mathcal{A}}
\newcommand{\ie}{i.e.}
\newcommand{\df}{{\mathrm{d}f}}
\newcommand{\rmi}{\mathrm{i}}
\newcommand{\rmd}{\mathrm{d}}
\newcommand{\rme}{\mathrm{e}}
\newcommand{\dt}{{\mathrm{d}t}}
\newcommand{\pj}{\partial_j}
\newcommand{\pk}{\partial_k}
\newcommand{\psifl}{\Psi(f; {\bm \lambda})}
\newcommand{\hp}{h_+(t)}
\newcommand{\hc}{h_\times(t)}
\newcommand{\Fp}{F_+}
\newcommand{\Fc}{F_\times}
\newcommand{\Ylm}{Y_{\ell m}^{-2}}
\def\no{\nonumber \\ & \quad}
\def\noQ{\nonumber \\}
\newcommand{\mc}{M_c}
\newcommand{\vek}[1]{\boldsymbol{#1}}
\newcommand{\vdag}{(v)^\dagger}
\newcommand{\btheta}{{\bm \theta}}
\newcommand{\pa}{\partial_a}
\newcommand{\pb}{\partial_b}
\newcommand{\Psieff}{\Psi_{\rm eff}}
\newcommand{\Aeff}{A_{\rm eff}}
\newcommand{\deff}{d_{\rm eff}}
\newcommand{\corr}{\mathcal{C}}
\newcommand{\bvthat}{\hat{\mbox{\boldmath $\vt$}}}
\newcommand{\bvt}{\mbox{\boldmath $\vt$}}

\newcommand{\comment}[1]{{\textsf{#1}}}
\newcommand{\ajith}[1]{\textcolor{magenta}{\textit{Ajith: #1}}}
\newcommand{\sukanta}[1]{\textcolor{blue}{\textit{Sukanta: #1}}}

\newcommand{\AEIHann}{Max-Planck-Institut f\"ur Gravitationsphysik 
(Albert-Einstein-Institut) and Leibniz Universit\"at Hannover, 
Callinstr.~38, 30167~Hannover, Germany}
\newcommand{\WSU}{Department of Physics \& Astronomy, Washington State University,
1245 Webster, Pullman, WA 99164-2814, U.S.A.}
\newcommand{\LIGOCaltech}{LIGO Laboratory, California Institute of Technology, 
Pasadena, CA 91125, U.S.A.}
\newcommand{\TAPIR}{Theoretical Astrophysics, California Institute of Technology, 
Pasadena, CA 91125, U.S.A.}

\title{Estimating the parameters of non-spinning binary black holes using 
ground-based gravitational-wave detectors: Statistical errors}
\preprint{LIGO-P0900002}

\author{P.~Ajith}
\email{ajith@caltech.edu}
\affiliation{\AEIHann}
\affiliation{\LIGOCaltech}
\affiliation{\TAPIR}

\author{Sukanta Bose}
\email{sukanta@wsu.edu}
\affiliation{\WSU}
\affiliation{\AEIHann}


\begin{abstract}

We assess the statistical errors in estimating the parameters of non-spinning 
black-hole binaries using ground-based gravitational-wave detectors. While 
past assessments were based on partial information provided by only the inspiral
and / or ring-down pieces of the coalescence signal, the recent progress in analytical and 
numerical relativity enables us to make more accurate projections using ``complete'' 
inspiral-merger-ringdown waveforms. We employ the Fisher information-matrix 
formalism to estimate how accurately the source parameters will be measurable 
using a single interferometric detector as well as a network of interferometers. 
Those estimates are further vetted by full-fledged Monte-Carlo simulations. We find that the 
parameter accuracies of the complete waveform are, in general, significantly better than 
those of just the inspiral waveform in the case of binaries with total mass $M  \gtrsim 20 M_\odot$. 
In particular, for the case of the Advanced LIGO detector, parameter estimation
is the most accurate in the $M=100-200 M_\odot$ range. For an $M=100M_\odot$ system, the 
errors in measuring the total mass and the symmetric mass-ratio are reduced by 
an order of magnitude or more compared to inspiral waveforms. Furthermore, for binaries 
located at a fixed luminosity distance $d_L$, and observed with the Advanced LIGO--Advanced 
Virgo network, the sky-position error is expected to vary widely across the sky: For 
$M=100M_\odot$ systems at $d_L=1$Gpc, this variation ranges mostly from about a hundredth 
of a square-degree to about a square-degree, with an average value of nearly a tenth of a 
square-degree. This is more than forty times better than the average sky-position accuracy 
of inspiral waveforms at this mass-range. For the mass parameters as well as the sky-position, 
this improvement in accuracy is due partly to the increased signal-to-noise ratio and partly to the 
information about these parameters harnessed through the post-inspiral phases of the waveform. The 
error in estimating $d_L$ is dominated by the error in measuring the wave's polarization and is 
roughly $43\%$ for low-mass ($M\sim20M_\odot$) binaries and about $23\%$ for high-mass 
($M\sim100M_\odot$) binaries located at $d_L=1$Gpc.

\end{abstract}
\maketitle

\section{Introduction}

Astrophysical black holes (BHs) are typically classified into three groups: stellar-mass 
BHs (with a mass of approximately 3 --- $30 M_\odot$), [super]massive BHs ($\sim10^4$ --- 
$10^{10} M_\odot$) and intermediate-mass (IM) BHs ($\sim30$ --- $10^4 M_\odot$). There is 
strong observational evidence for the existence of both stellar-mass and supermassive BHs. 
The existence of stellar-mass BHs, which are the end products of stellar evolution, has 
been primarily inferred from observations of X-ray binaries that allow us to estimate the 
mass of the compact object through measurements of the orbital period and the maximum 
line-of-sight Doppler velocity of the companion star~\cite{RNarayan:BHReview}. The mechanism 
for producing supermassive BHs is less certain but the acceleration of gas disks in the 
bulges of nearly all local massive galaxies point to their existence there~\cite{KormendyAndRichstone:1995}.
Even more convincingly, the observations of stellar proper motion in the center of the Milky 
Way have confirmed the presence of a supermassive BH~\cite{RSchoedel:2002Nature}. On the 
other hand, the  observational evidence for IMBHs is only suggestive. The main hint comes 
from the observations of ultraluminous X-ray sources, combined with the fact that several 
globular clusters show evidence for an excess of dark matter in their cores~\cite{MillerColbert:2004}. 

According to hierarchical galaxy-merger models,  [super]massive BH binaries should 
form frequently, and should be common in the cores of galaxies. There is at least one 
piece of clear evidence for the existence of a supermassive BH binary, namely, 
the X-ray active binary black hole (BBH) at the 
center of the galaxy NGC~6240, which is expected to coalesce in Hubble time~\cite{Komossa:2002tn}. 
There is also growing observational evidence for the existence of many other [super]massive 
BBHs~\cite{Ballo:2003ww,Guainazzi:2004ky,Evans:2007nq,Bianchi:2008un}. Despite the lack of 
any observational evidence for stellar-mass/intermediate-mass BH binaries, different 
mechanisms to form these binaries have been proposed in the literature (see, for e.g.,
\cite{lrr-PostnovYungelson,AmaroSeoane:2006py,Fregeau:2006yz,Mandel:2007hi}).

Coalescing BH binaries are among the most promising sources of gravitational
waves (GWs) for the ground-based interferometric detectors. What makes them
extremely interesting is that their gravitational waveforms can be 
accurately modelled and well parametrized by combining a variety of analytical and 
numerical approaches to General Relativity. To wit, the gravitational waveforms 
from the \emph{inspiral} stage of the binary can be accurately computed by the post-Newtonian (PN)
approximation to General Relativity, while those from the \emph{ring down} stage can be 
computed using BH perturbation theory. The recent breakthrough~\cite{Pretorius:2005gq,
Campanelli:2005dd,Baker05a} in numerical relativity has made it possible to compute 
accurate gravitational waveforms from the hitherto unknown \emph{merger} stage as 
well~\cite{Pretorius:2005gq,Campanelli:2005dd,Baker05a,Herrmann2006,
Sperhake2006,Bruegmann:2006at,Thornburg-etal-2007a,Etienne:2007hr}.  

Concomitant with that breakthrough has been the notable progress in GW instrumentation.
The Initial LIGO (LIGOI)~\cite{Sigg-LIGOstatus-2008} detectors have completed their first science run at 
design sensitivity. The Virgo detector~\cite{VirgoStatus-GWDAW2008} ran concurrently with 
LIGO for part of that run. Currently, both observatories are undergoing commissioning work 
with the target of achieving second-generation sensitivities over the next several years, 
to usher us into the era of Advanced LIGO (AdvLIGO)~\cite{AdLigoUrl} and Advanced Virgo (AdvVirgo).
Also, an intermediate, enhanced stage of LIGO, called Enhanced LIGO (EnhLIGO), is expected to 
be operational this year. 

In the absence of any observational evidence of stellar-mass/intermediate-mass
BH binaries, the rate of binary coalescence events is estimated by population synthesis studies. 
Plausible rate estimates for stellar-mass BH coalescences detectable by LIGOI / EnhLIGO / AdvLIGO 
detectors range from $7\times10^{-4}\,/\,7\times10^{-3}\,/\,2$ per year to $2\,/\,20\,/\,4000$ per 
year with a likely rate estimate of around $0.01\,/\,0.1\,/\,30$ per year~\cite{O'Shaughnessy:2005qs}. 
For the case of IMBH binaries, the plausible rates for LIGOI / AdvLIGO detectors
are $10^{-4} \,/\,0.1$ per year~\cite{Fregeau:2006yz}. Similarly, for the case of stellar-mass
BHs merging with IMBHs (the so-called intermediate-mass-ratio inspirals), plausible
event rates for LIGOI / AdvLIGO are $10^{-3}\,/\,10$ per year~\cite{Mandel:2007hi}~\footnote{It should 
be noted that these assessments take into account only the inspiral stage (for the case
of stellar-mass and intermediate-mass-ratio binaries) or ring-down stage (for the case of IMBH binaries)
of the binary coalescence. The event rates are likely to be higher for a search using inspiral-merger-
ring down templates. See, for example, Fig.~14 of \cite{Ajith:2007kx} for a comparison of the 
sensitivities of searches employing different templates.}. 
A network of interferometric detectors involving LIGO, Virgo, and perhaps others, such as 
GEO600~\cite{Grote-GEOstatus-2008}, and TAMA~\cite{Takahashi-TAMAstatus-2008}, will be able to 
extract a host of physical parameters of those sources, complementing other detectors probing 
their electromagnetic characteristics.

Indeed, some of the BBH mergers, e.g., triggered by the mergers of galaxies/stellar 
clusters harboring supermassive/intermediate-mass BHs, are likely to have electromagnetic 
(EM) counterparts. To associate an EM event with a GW signal from such a merger, and vice 
versa, one needs to be able to locate the GW source with a high enough accuracy so that the 
number of star clusters or galaxies in the sky-position error box is sufficiently small. As 
argued in Ref. \cite{Holz:2005df}, even arc-minute resolution can make such associations quite 
feasible. Whereas the GW observations are expected 
to provide more accurate distance measurements than their EM counterpart, the latter will locate 
the sources in the sky with far greater resolution than the former. This complementarity was explored 
in Ref.~\cite{Schutz86} to argue that by combining  GW and electromagnetic observations it 
should be possible to constrain the values of certain cosmological parameters.
In particular, using the distance-redshift relation from many BBH ``standard sirens'', such 
multi-messenger observations can put interesting constraints on the equation of state of the dark 
energy~\cite{HolzHugh05,Arun:2007hu}. Supermassive BH binaries are also excellent test beds for 
``strong-field'' predictions of General Relativity (see, e.g.,~\cite{Arun:2006yw,Arun:2006hn}). 
Also, GW observations of BBH coalescences can be used to test theoretical predictions such as 
the ``no-hair'' theorem~\cite{Ryan:1995wh}. The effectiveness of these and other applications 
depends on the accuracy with which we can estimate the parameters of the binary, which includes 
the component masses, distance, orientation, and sky location. 

In this work, we study the effect of detector noise in limiting the accuracy with which parameters 
of a BBH system can be determined with the present and planned earth-based laser interferometers.
In the past, in the absence of complete coalescence waveforms arising from numerical 
relativity, parameter estimation studies were constrained to address this question only 
for the inspiral/ring-down pieces of the signal present in the band of a detector~\cite{Markovic:1993cr,
Cutler:1994ys,Balasubramanian:1995ff,Balasubramanian:1995bm,Balasubramanian:1997qz,Nicholson:1997qh,
Jaranowski:1996hs,Poisson:1995ef,Arun:2004hn,LunaSintes06}. 
Here we extend those studies to estimate how the astrophysical quest for characterizing
such systems benefits from the knowledge of the complete coherent signal, comprising some 
or all of the inspiral, merger, and ringdown pieces, that lies in a detector's observational 
band. Improvements in the accuracy of BBH parameter measurements might be 
expected owing to the increased signal-to-noise ratio (SNR) arising from the inclusion of 
the post-inspiral pieces. A second avenue toward parameter accuracy improvements can also 
arise, for some parameters, from the breaking of some parameter degeneracies that the extra information carried 
by the GW phasing of those pieces might offer. We employ the phenomenological 
inspiral-merger-ringdown waveforms proposed in Refs.~\cite{Ajith:2007kx,Ajith:2007qp,Ajith:2007xh} 
to explore these possibilities.~\footnote{A similar study using the effective-one-body-numerical-relativity 
waveforms~\cite{Buonanno:2007pf,Pan:2007nw,Damour:2007vq,Damour:2007yf,Damour:2008te} is being 
pursued as well~\cite{SathyaCommunication}.} The {\em systematic} errors that might arise in 
observations using these ``complete''~\footnote{Throughout this paper,
we refer to the waveforms modelling all the three (inspiral, merger and ringdown)
stages of BBH coalescence as ``complete'' waveforms.} BBH coalescence 
templates are studied in Ref.~\cite{AjithEtAlParamEstimSyst}.

To estimate the parameter errors, we adopt a two-pronged approach. One of these is of 
obtaining the Fisher information matrix and then inverting it to derive the parameter 
error variance-covariance matrix~\cite{Helstrom}. The elements of this matrix are then 
used to obtain the lower bound on the parameter estimator errors~\cite{Cramer46,Rao45}. This 
approach is employed here, in spite of its known limitations~\cite{Balasubramanian:1995ff,
Balasubramanian:1995bm,Vallisneri:2007ev}, since it has been studied extensively in the 
community and allows for a fair comparison of our results with those given in the literature. 
However, since by its very design, this bound may not be respected for signals with a low 
SNR (as first demonstrated by Refs.~\cite{Balasubramanian:1995ff,Balasubramanian:1995bm}), we also assess 
estimator errors through Monte Carlo studies. 
For the parameter ranges considered here, the latter approach corroborates the findings of the former, with a 
few notable exceptions arising from parameter space boundaries, where the Monte Carlo 
estimates reflect better the results expected from real-data searches.

In addition to addressing the primary question on how large the parameter errors are, we 
also study their behavior across the BBH parameter space. We study how the various estimator 
errors scale with the mass parameters themselves. How much improvement do the complete waveforms 
impart to the determination of the sky-position of BBHs in 
multi-detector searches? How does the sky-position accuracy change with the BBH mass parameters?
A summary of our results is as follows: First, we find that the 
parameter-estimation accuracies using the complete waveforms are, in general, 
significantly better than those using only their inspiral phases in the case of
BBHs with a total mass $M\equiv (m_1 + m_2) \gtrsim 20 M_\odot$, where 
$m_{1,2}$ are the component masses, at least for mass-ratios between 0.25 and unity. The 
observed trend suggests that this improvement can be expected for somewhat lower mass-ratios 
as well. Second, for BBHs at a fixed effective distance and 
$M \gtrsim 10 M_\odot$ whereas the fractional errors in the 
two mass parameters, $M$ and $\eta \equiv m_1m_2/M^2$, scale mostly monotonically with 
$M$ for the inspiral-only waveforms, they do not display that property for the complete 
waveforms. In the latter case, they instead exhibit a distinct minimum, whose location is 
determined by $M$, $\eta$, and the detector's noise power spectral-density (PSD). Third, 
owing to the use of complete {\it vis \`{a} vis} inspiral-only waveforms the sky-position 
accuracy improves by factors of many. We also show that for the complete waveforms alone, 
the sky-position accuracy mostly degrades with increasing total-mass when the SNR is kept 
fixed. This is primarily caused by a similar 
degradation in the estimation accuracy of the signal's times of arrival at the different 
detectors in a network. This deterioration in accuracy, while not monotonic in $M$ at finer 
scales, is broadly so at large scales, and is caused by the reduction in the number of in-band 
wave cycles.

More specifically, for Advanced LIGO, the estimation of the total mass, the symmetric
mass-ratio, and the effective distance $d_{\rm eff}$ is the most accurate in the $M=100-200 M_\odot$ 
range. (For other detectors, that mass range is somewhat different since it is partly determined 
by their noise PSDs.) For such systems, the reduction of errors in parameter estimates is 
by an order-of-magnitude or more due to the inclusion of the post-inspiral phases. The improvement 
is mainly due to the expected increase in SNR arising from the inclusion 
of those phases. This expectation, which is based on the assumed Gaussianity and stationarity of 
detector noise, must be tempered by the observation that the amount of increase in SNR 
can be less in real data.

We also observe that for a fixed SNR, the inclusion of the post-inspiral phases improves the 
accuracy of $M$ and $\eta$ for a wide range of masses much more 
(by several times) than that of the chirp mass $\mc$. This is due to the fact that the 
inclusion of those phases helps in breaking the degeneracy between those 
two parameters ($M$ and $\eta$) known to exist in the inspiral waveform. 

For a fixed SNR, the estimation of the luminosity distance for low-mass systems
shows negligible change by the inclusion of the post-inspiral phases. This is 
due to its strong covariance with the polarization and the orbital inclination 
angles of the binary, which is mostly unaltered by the inclusion of the 
post-inspiral phases. Also, for a fixed SNR, the luminosity distance estimate 
deteriorates with increasing $M$, for reasons discussed below. On the other hand, 
for a fixed luminosity distance, the error in its estimate initially improves 
with increasing $M$, due to the increase in SNR, before degrading eventually 
owing to the decreasing number of in-band wave cycles.

Before moving on, we wish to point out some limitations of the present work. First, 
this study considers only the dominant harmonic of non-spinning BBH waveforms. 
Astrophysical BHs are expected to have spin, and including spin effects 
can change the estimation of different BBH parameters~\cite{BBW2004}. Whereas on the 
one hand previous calculations have shown that the parameter-estimation accuracies generally 
deteriorate upon the inclusion of spin-orbit and spin-spin couplings~\cite{BBW2004,Cutler:1994ys}, 
on the other hand the inclusion of spin-induced precession 
in the waveform model can improve the parameter estimation~\cite{Vecchio04,LangHughes06}. 
Also, it has been noted in various studies that including the higher harmonics can significantly 
increase the parameter-estimation accuracies~\cite{Moore:1999zw,Sintes:1999cg,ChrisAnand06b,
Arun:2007hu,Trias:2008pu,Porter:2008kn}. 
So, while the results presented in this paper may not be too far from the realistic 
case, we stress that a rigorous statement on the parameter-estimation accuracies should
consider these effects as well. Moreover, neglecting spins and higher harmonics in the waveform 
models can result in significant amount of systematic errors in estimating various parameters. 
These systematic errors are out of the scope of this paper. A preliminary investigation of 
this is presented in Ref.~\cite{AjithEtAlParamEstimSyst}.  

This paper is organized as follows: Sec.~\ref{sec:PhenWave} briefly introduces the main 
aspects of the search for binary black holes. In particular, Sec.~\ref{sec:PhenWaveDetect} 
reviews the phenomenological inspiral-merger-ring down waveform templates proposed in 
Refs.~\cite{Ajith:2007kx,Ajith:2007qp,Ajith:2007xh}, while Sec.~\ref{sec:ParamEstimTheory} 
provides a brief introduction towards the statistical theory of parameter estimation. In 
Sec.~\ref{sec:ParamEstimSingleDet}, we present the results of our calculations in the case of 
a search using a single interferometric detector. This section discusses the results from the analytical 
calculations using the Fisher-matrix formalism as well as the numerical Monte-Carlo simulations. 
Results from the calculations in the case of a network of detectors are discussed
in Sec.~\ref{sec:ParamEstimMultDet}, while Sec.~\ref{sec:Summary} summarizes the main results
and provides a discussion of the possible astrophysical implications of this work. 

\section{Gravitational wave observations of non-spinning binary black holes}
\label{sec:PhenWave}

In General Relativity, the gravitational-wave strain at any point in space
can be expanded in terms of its two linear polarization components
$\hp$ and $\hc$ or the two related circular polarization components,
\be
\h (t) \equiv \hp - \rmi \hc = A(t)\, e ^{\rmi \varphi(t)}
\ee
and its complex conjugate, with $\varphi(t)$ and $A(t)$ denoting the wave's phase and amplitude.
Generally, the GW emitted by a coalescing binary has multiple harmonics. In this work, we 
limit our study to only the dominant harmonic's contributions to $\varphi(t)$ and $A(t)$. 
Then the GW strain $h(t)$ in a detector is the linear combination 
of the two polarization components, $h(t) = \Fp \hp + \Fc \hc$, with the
detector's antenna-pattern functions given as: 
\ber
F_+(\theta,\phi,\psi) & = & -\frac{1}{2} (1+\cos^2 \theta) \cos 2 \phi~\cos 2\psi \nonumber \\
& - & \cos \theta ~ \sin 2\phi ~ \sin 2\psi, \nonumber \\
F_\times(\theta,\phi,\psi) & = & \frac{1}{2} (1+\cos^2 \theta) \cos 2\phi~\sin 2\psi \nonumber \\
& - & \cos\theta ~\sin2\phi ~ \cos 2\psi. 
\eer
Above, $\theta$ and $\phi$ are the polar and azimuthal angles specifying the location of the 
source in the sky in the detector frame and $\psi$ is the polarization angle.

The two polarization components of the BBH signals are sinusoids with varying 
amplitude and frequency, and have phases $\pi/2$ radians apart relative to each
other. Consequently, their GW signal in a detector can be written as:
\be
h(t) = C \,  A(t) \,\cos [\varphi(t) + \varphi_0],
\label{eq:hOfTDomMode}
\ee 
where the amplitude coefficient $C$ and phase $\varphi_0$ can be assumed to be 
constant for signals lasting for a duration (up to several minutes) much shorter 
than Earth's rotational time-scale:
\ber
C &=& \frac{1}{2} \sqrt{ (1+\cos^2 \iota)^2 \Fp^2 + 4 \cos^2 \iota \Fc^2 }, \nonumber \\
\varphi_0 &=& \tan^{-1} \left[ \frac{2 \Fc \cos \iota }{\Fp (1+\cos^2 \iota)} \right]\,.
\label{eq:CAndPhi0DomMode}
\eer
Above, $\iota$ is angle of inclination of the orbit to the line of sight.

\begin{figure}[tb]
\centering
\includegraphics[width=3.5in]{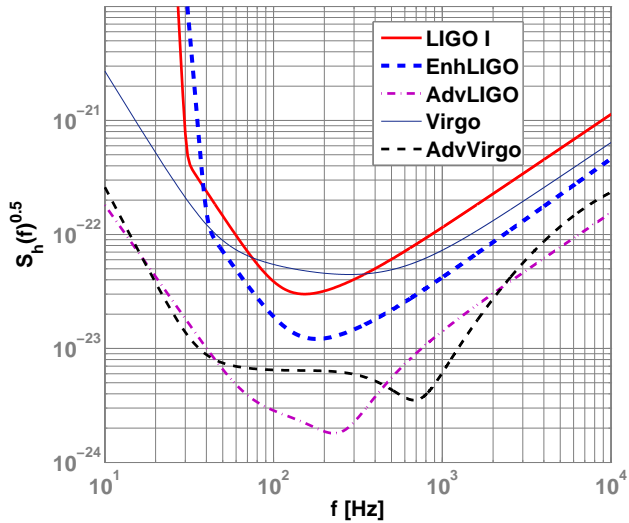}
\caption{Noise amplitude spectrum ($\sqrt{S_h(f)}$) of different 
detectors considered in this paper.}
\label{fig:NoisePSD}
\end{figure}

\subsection{Detecting non-spinning binary black holes} 
\label{sec:PhenWaveDetect}

The GW signal's phase $\varphi(t)$ and amplitude $A(t)$ are functions of the 
physical parameters of the binary, such as the component masses and the spins. 
Detecting a signal requires analyzing interferometric data, which are noisy. 
Defining a search strategy, therefore, necessitates the modelling of this noise,
which we take here to be zero-mean Gaussian and stationary:
\begin{mathletters}%
\label{noise}
\bea
{\overline{n(t)}} &=& 0, \label{noisea}\\
{\overline{{\tilde n}^*(f){\tilde n}(f')}} &=& \frac{1}{2}S_{h}(f)\, \delta(f-f')\ , \label{noiseb}
\eea
\end{mathletters}%
with the over-bar denoting the ensemble average
and the tilde denoting the Fourier transform, 
\be \label{FTdef}
{\tilde n}(f) = \int_{-\infty}^{\infty} n(t)\, e^{-2\pi \rmi f t} \,\dt \,.
\ee
Above, $S_{h}(f)$ is the Fourier transform of the auto-covariance of the 
detector noise and is termed as its (one-sided) power spectral-density.
We also assume the noise to be additive. This implies that 
when a signal is present in the data $x(t)$, then 
\be
x(t) = h(t) + n(t) \,.
\ee
The noise covariance Eq. (\ref{noiseb}) introduces the following inner-product in the 
function space of signals:
\be\label{innerprod}
\langle a ,\>b \rangle = 4 \Re \int_{0}^{\infty} \! \df\>
{\tilde{a}^* (f) \,\tilde{b}(f) \over S_{h}(f)} \ \ , 
\ee
where $\tilde{a}(f)$ and $\tilde{b}(f)$ are the Fourier
transforms of $a(t)$ and $b(t)$, respectively.

Under the above assumptions about the characteristics of detector noise,
the Neyman-Pearson criterion \cite{Helstrom} leads to an optimal search statistic,
which when maximized over the amplitude coefficient $C$, is the cross-correlation 
of the data with a normalized template,
\be\label{crosscor}
\rho \equiv \langle \hat{h},x\rangle \ \ ,
\ee
where the normalized template is $\hat{\tilde h}(f) \equiv {\tilde h}(f)/\sqrt{\langle h,\>h \rangle} $. 
In a ``blind'' search in detector data, where none of the binary's parameters are known \emph{a priori}, 
the search for a GW signal requires maximizing $\rho$ over a ``bank'' of templates (see, for 
e.g.,~\cite{Cokelaer:2007kx}) corresponding to different values of those physical parameters. Apart from the 
physical parameters, the waveform also depends on the (unknown) initial phase $\varphi_0$ and the 
time of arrival $t_0$. Maximization over the initial phase $\varphi_0$ is effected by using two orthogonal 
templates for each combination of the physical parameters~\cite{schutz-91}, and the maximization over 
$t_0$ is attained efficiently with the help of the Fast Fourier Transform (FFT) algorithms~\cite{NRecipes}. 

Since the cross correlation between the data and the template can be most efficiently
computed in the Fourier domain by using the FFT, waveform templates in the Fourier domain 
are computationally cheaper. Reference~\cite{Ajith:2007kx} proposed a family of analytical Fourier 
domain templates for BBH waveforms of the form:
\be
{\tilde h}(f) \equiv {A}_{\rm eff}(f) \, e^{\rmi\Psi_{\rm eff}(f)},
\label{eq:phenWave}
\ee
where the effective amplitude and phase are expressed as:
\begin{widetext}
\ber
{A_{\rm eff}}(f) &\equiv& \frac{M^{5/6}}{\deff\,\pi^{2/3}}\sqrt{\frac{5\,\eta}{24}}\,\fmerg^{-7/6}
\left\{ \begin{array}{ll}
\left(f/\fmerg\right)^{-7/6}   & \textrm{if $f < \fmerg$}\\
\left(f/\fmerg\right)^{-2/3}   & \textrm{if $\fmerg \leq f < \fring$}\\
w \, {\cal L}(f,\fring,\sigma) & \textrm{if $\fring \leq f < \fcut$,}\\
\end{array} \right. \nonumber \\
\Psi_{\rm eff}(f) &\equiv& 2 \pi f t_0 + \varphi_0 + \frac{1}{\eta}\,\sum_{k=0}^{7} 
(x_k\,\eta^2 + y_k\,\eta + z_k) \,(\pi M f)^{(k-5)/3}\,.
\label{eq:phenWaveAmpAndPhase}
\eer
\end{widetext}
In the above expressions,
\be
{\cal L}(f,\fring,\sigma) \equiv \left(\frac{1}{2 \pi}\right) 
\frac{\sigma}{(f-\fring)^2+\sigma^2/4}\,
\ee 
is a Lorentzian function that has a  width $\sigma$, and that is centered
around the frequency $\fring$. The normalization constant,
$w \equiv \frac{\pi \sigma}{2} \left(\frac{f_{\rm ring}}
{f_{\rm merg}}\right)^{-2/3}$, is chosen so as to make 
${A}_{\rm eff}(f)$ continuous across the ``transition'' frequency $f_{\rm ring}$. 
The parameter $f_{\rm merg}$ is the frequency at which the power-law changes 
from $f^{-7/6}$ to $f^{-2/3}$. The \emph{effective distance} to the binary is denoted
by $\deff$, which is related to the luminosity distance $d_L$ by $\deff = d_L/C$.
The phenomenological parameters $\fmerg, \fring, \sigma$ and $\fcut$ 
are given in terms of the total mass $M$ and symmetric mass-ratio $\eta$ of the 
binary as
\ber
\pi M \fmerg &=&  a_0 \, \eta^2 + b_0 \, \eta + c_0  \,, \nonumber \\
\pi M \fring &= & a_1 \, \eta^2 + b_1 \, \eta + c_1  \,, \nonumber \\
\pi M \sigma &= & a_2 \, \eta^2 + b_2 \, \eta + c_2  \,, \nonumber \\
\pi M \fcut  &= & a_3 \, \eta^2 + b_3 \, \eta + c_3. 
\label{eq:ampParams}
\eer
The coefficients $a_j, b_j, c_j,~j=0...3$ and $x_k,y_k,z_k,~k=0,2,3,4,6,7$ are 
tabulated in Table I of Ref.~\cite{Ajith:2007xh}.

\subsection{Measuring binary black hole parameters} 
\label{sec:ParamEstimTheory}

\begin{figure*}[tb]
\centering
\includegraphics[width=6.1in]{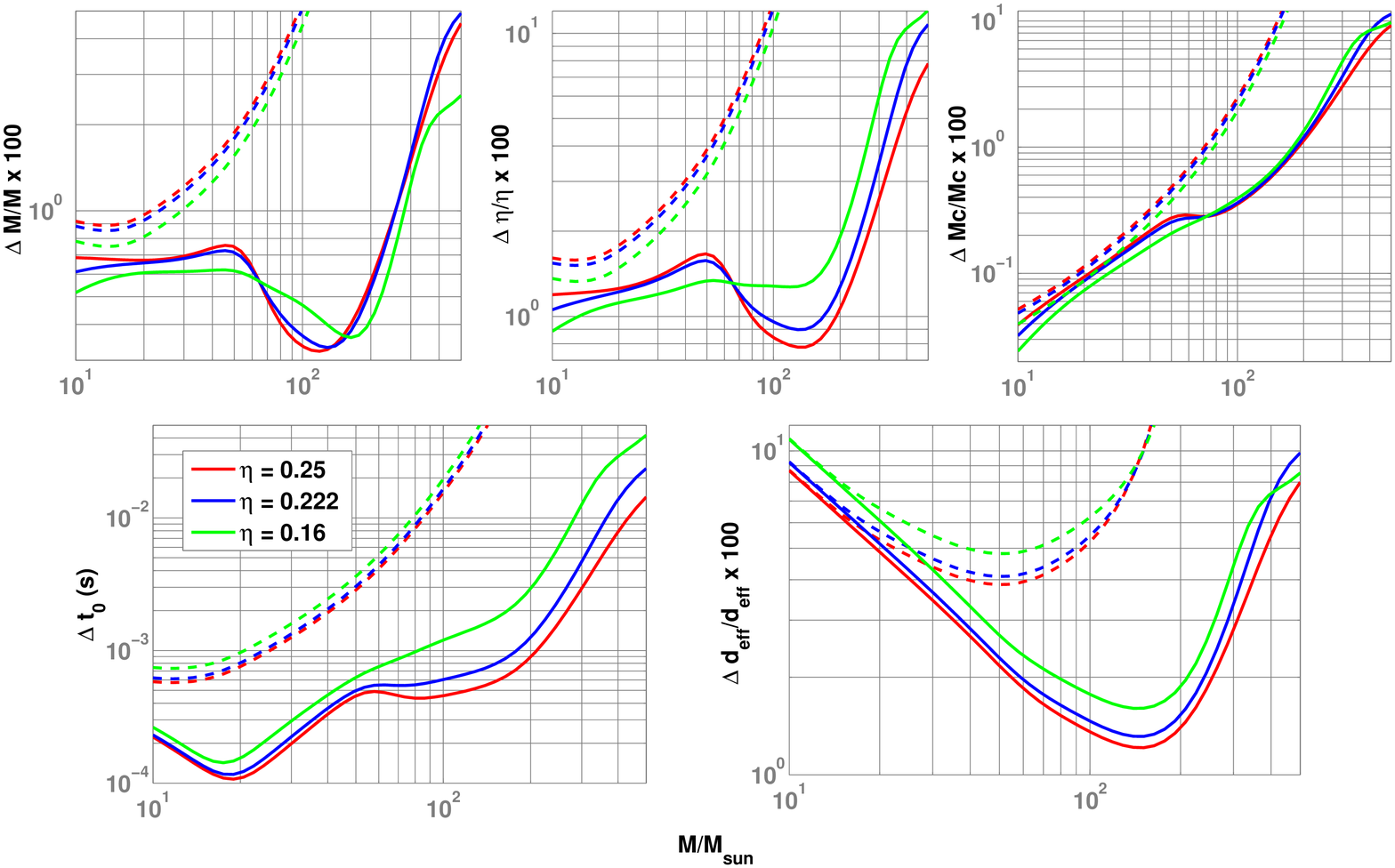}
\caption{Errors in estimating the total mass $M$ (top left), symmetric mass ratio 
$\eta$ (top middle), chirp mass $\mc$ (top right), time of arrival $t_0$ (bottom 
left) and effective distance $d_{\rm eff}$ (bottom right) in the case of Advanced LIGO noise spectrum,
plotted against the total mass of the binary. The errors of $M, \eta, \mc$ and 
$d_{\rm eff}$ are in percentage and the errors of $t_0$ are in seconds. The value of
the symmetric mass-ratio $\eta$ is shown in legends. The solid lines correspond to a 
search using complete BBH templates and the dashed lines correspond to a search using 
3.5PN-accurate post-Newtonian templates in the SPA, truncated at the Schwarzschild ISCO. The 
binary is placed optimally oriented at an effective distance of 1 Gpc.}
\label{fig:FishMatErrorsFixdDist}
\end{figure*}

\begin{figure*}[tb]
\centering
\includegraphics[width=6.1in]{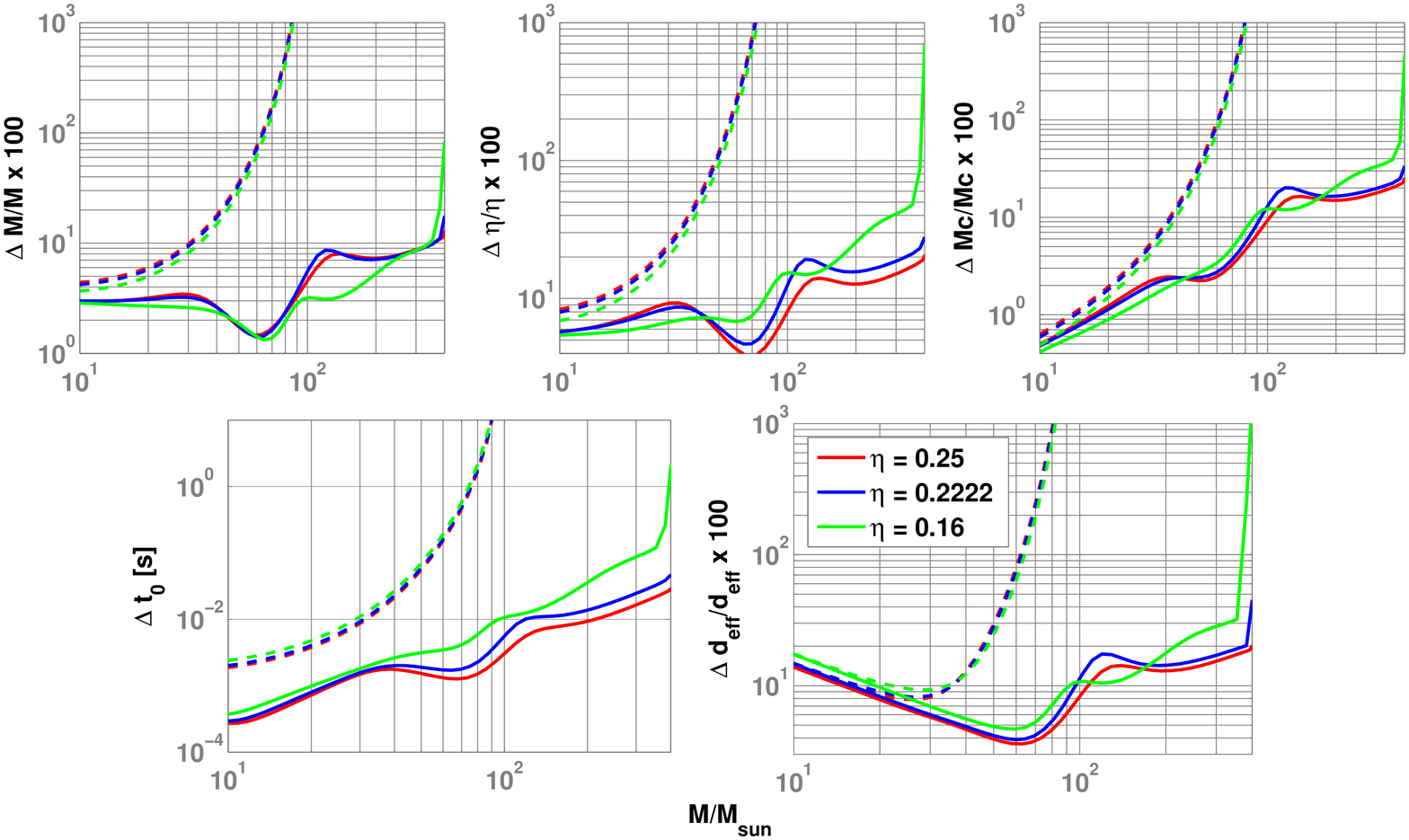}
\caption{Same as in Fig.~\ref{fig:FishMatErrorsFixdDist} except that the binary
is placed at an effective distance of 100 Mpc and the noise PSD corresponds to 
that of Initial LIGO.}
\label{fig:FishMatErrorsFixdDistLIGOI}
\end{figure*}

\begin{figure*}[tb]
\centering
\includegraphics[width=6.1in]{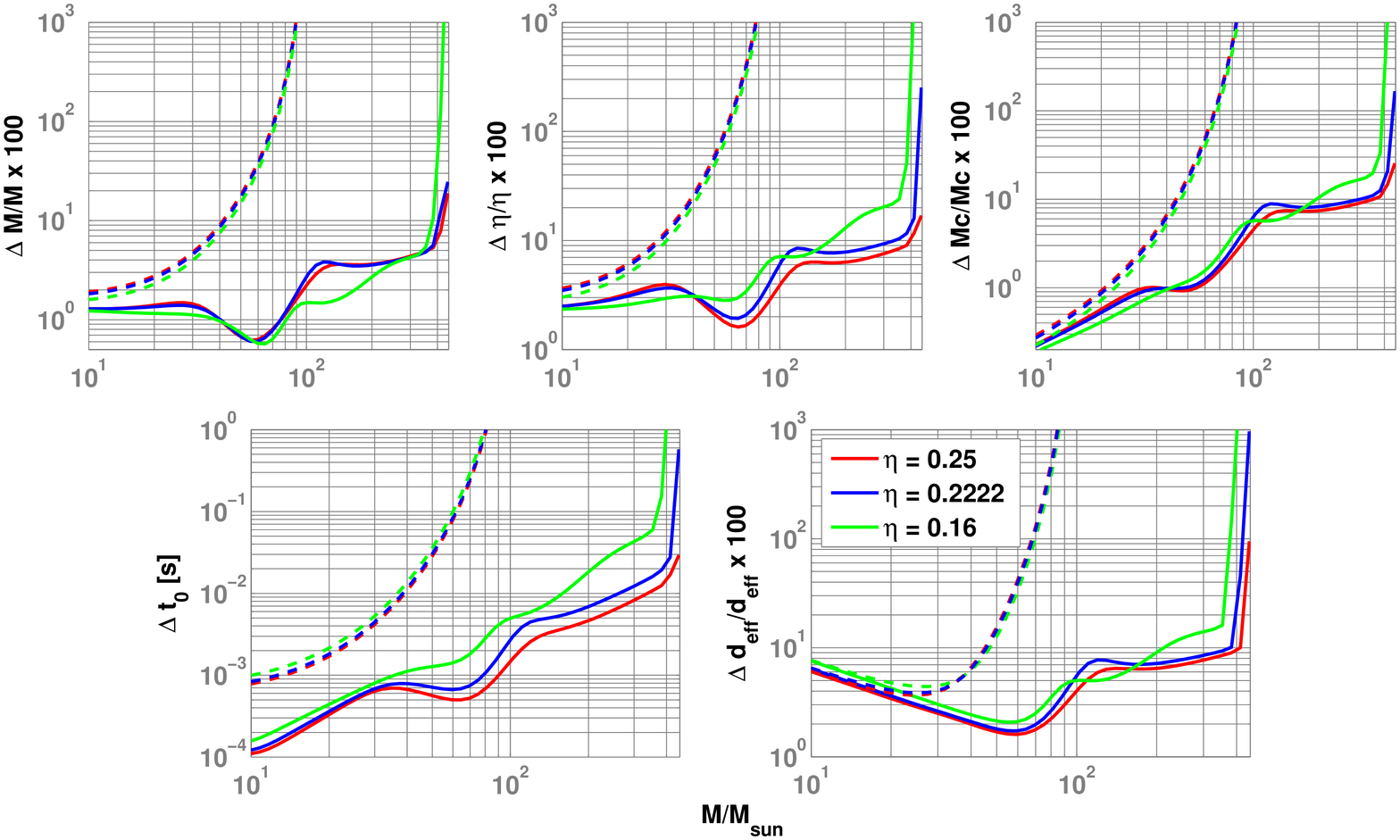}
\caption{Same as in Fig.~\ref{fig:FishMatErrorsFixdDist} except that the binary
is placed at an effective distance of 100 Mpc and the noise PSD corresponds to 
that of Enhanced LIGO.}
\label{fig:FishMatErrorsFixdDistEnhLIGO}
\end{figure*}

To evaluate how effective the detectors will be in establishing the 
field of GW astronomy, especially, with the second-generation 
Earth-based interferometers scheduled to come online around 2014,
one needs to foremost assess how accurately they can measure 
the astrophysical properties of compact object binaries. That quest will be limited, 
on the one hand, by the accuracy with which the search templates can model actual 
gravitational waveforms, and, on the other hand, by the inherent statistical noise 
in the measurement process. The former issue is one of systematics, which will be 
discussed elsewhere (see, e.g., Ref.~\cite{AjithEtAlParamEstimSyst}). Here, we discuss 
the latter issue in more detail.

To determine how large the noise-limited errors can be in the measured values
of the signal parameter, we take those values to be the maximum likelihood 
estimators (MLEs). The discussion in the preceding section shows that a total 
of nine parameters characterize the non-spinning 
BBH coalescence signals considered here.
They are the total mass $M$, the symmetric mass-ratio $\eta$, the sky-position
angles $(\theta,\phi)$, the binary's orientation angles $(\psi,\iota)$,
the luminosity distance $d_L$, the initial (or some reference) phase 
$\varphi_0$, and the time of arrival (or some reference time) $t_0$. For 
computing the error estimates, we map them onto the components of the 
parameter vector, 
$\mbox{\boldmath $\vt$} \equiv \{\ln \cA, t_0, \varphi_0, \ln M, \ln \eta, \theta,\phi,\psi,\iota\}$,
where $\cA= \frac{M^{5/6}}{\deff \pi^{3/2}} \sqrt{\frac{5 \eta}{24}}$. 
Owing to noise, their MLEs, $\hat{\mbox{\boldmath $\vt$}}$, will expectedly 
fluctuate about the true values, i.e., $\bvthat = \bvt + \delta  \bvt$, 
where $\delta \mbox{$\vt^a$}$ is the random error in estimating the parameter 
$\mbox{$\vt^a$}$. The magnitude of these fluctuations can be quantified by
the elements of the variance-covariance matrix, 
$\gamma^{ab}=\ \overline{\delta \vt^a\,\delta \vt^b}$ \cite{Helstrom}.

A relation between the $\gamma^{ab}$ and the signal is available through
the Cramer-Rao inequality, which dictates that
\be
\parallel {\mbox{\boldmath $\gamma$}}\parallel ~\geq~ \parallel {
\mbox{\boldmath $\Gamma$}}\parallel^{-1} \ \ ,
\ee
where $\mbox{\boldmath $\Gamma$}$ is the Fisher information matrix:
\bea
\label{Fisher} \Gamma_{ab} &=& \sum_{I=1}^N
\left< \partial_a {\tilde h}^I(\mbox{\boldmath $\vt$}),
\partial_b {\tilde h}^I(\mbox{\boldmath $\vt$}) \right>_{(I)} \nonumber \\
&\equiv& \sum_{I=1}^N 4 \Re \int \df~
{\partial_a {\tilde h}^{I*}(f; \mbox{\boldmath $\vt$})~
\partial_b {h}^I(f; \mbox{\boldmath $\vt$})
\over S_{h}^I(f)},
\eea
where $I$ is the detector index and $\pa$ denotes taking partial derivative 
with respect to the parameter $\vt^a$. Therefore, 
$\Delta \vt^a \equiv \left(~\overline{\delta \vt^a\,\delta \vt^a}~\right)^{1/2} = \Gamma_{aa}^{-1/2}$ gives 
the lower bound on the root-mean-square (rms) error in estimating $\vt^a$. 
The two are equal in the limit of large SNR (see, e.g.,~\cite{Vallisneri:2007ev}).

The errors in the sky-position angles will be presented in terms of the
error in the measurement of the sky-position {\em solid angle}, defined as:
\be \label{angularRes}
\Delta\Omega = 2 \pi \sqrt{ (\Delta\cos\theta \, \Delta\phi)^2 -\left(~\overline{\delta\cos\theta~\delta\phi}~\right)^2} \,.
\ee
Each parameter-error, $\Delta\vt^a$, falls off inversely with 
SNR. Since the solid angle is two dimensional, its error falls off quadratically 
with SNR \cite{Helstrom,Cutler:1997ta,Rogan:2006nb}.

\section{Parameter estimation: Single-detector search}
\label{sec:ParamEstimSingleDet}

\begin{figure*}[tb]
\centering
\includegraphics[width=6.8in]{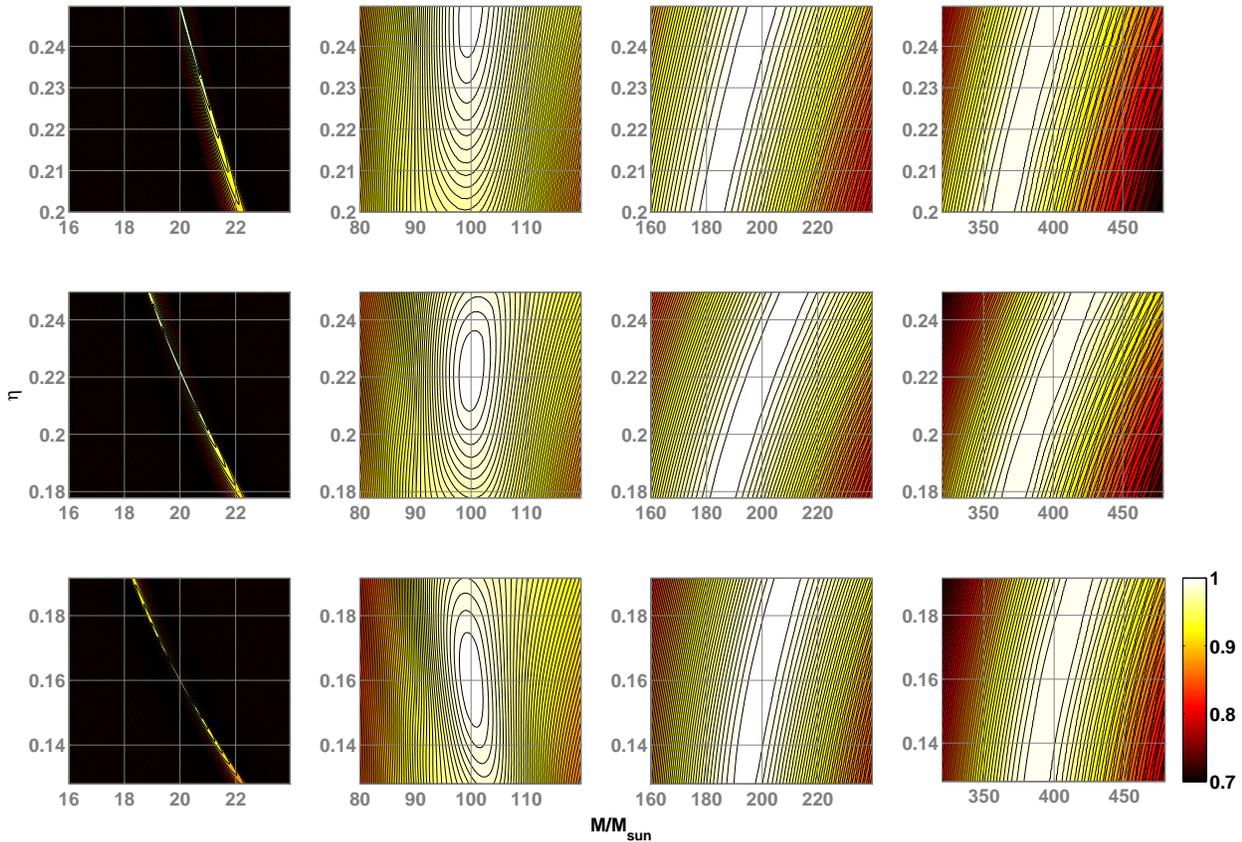}
\caption{The overlap function (i.e., the ambiguity function maximized over $t_0$ 
and $\varphi_0$) between waveforms constructed at different points in the parameter space. The 
horizontal axis reports the total mass $M$ of the binary while the vertical axis reports its
symmetric mass-ratio $\eta$. Each panel shows the overlap of different waveforms with one 
``target waveform''. The total mass of the target waveform is chosen to be 
$M = 20 M_\odot,~100 M_\odot,~200 M_\odot,~400 M_\odot$, respectively, for the 
four columns starting from the left. The symmetric mass-ratio of the target 
waveforms is chosen to be $\eta = 0.25,~0.222,~0.16$ in the top, middle 
and bottom rows, respectively.}
\label{fig:OverlapFnContourAdvLIGO}
\end{figure*}

\begin{figure*}[tb]
\centering
\includegraphics[width=6.1in]{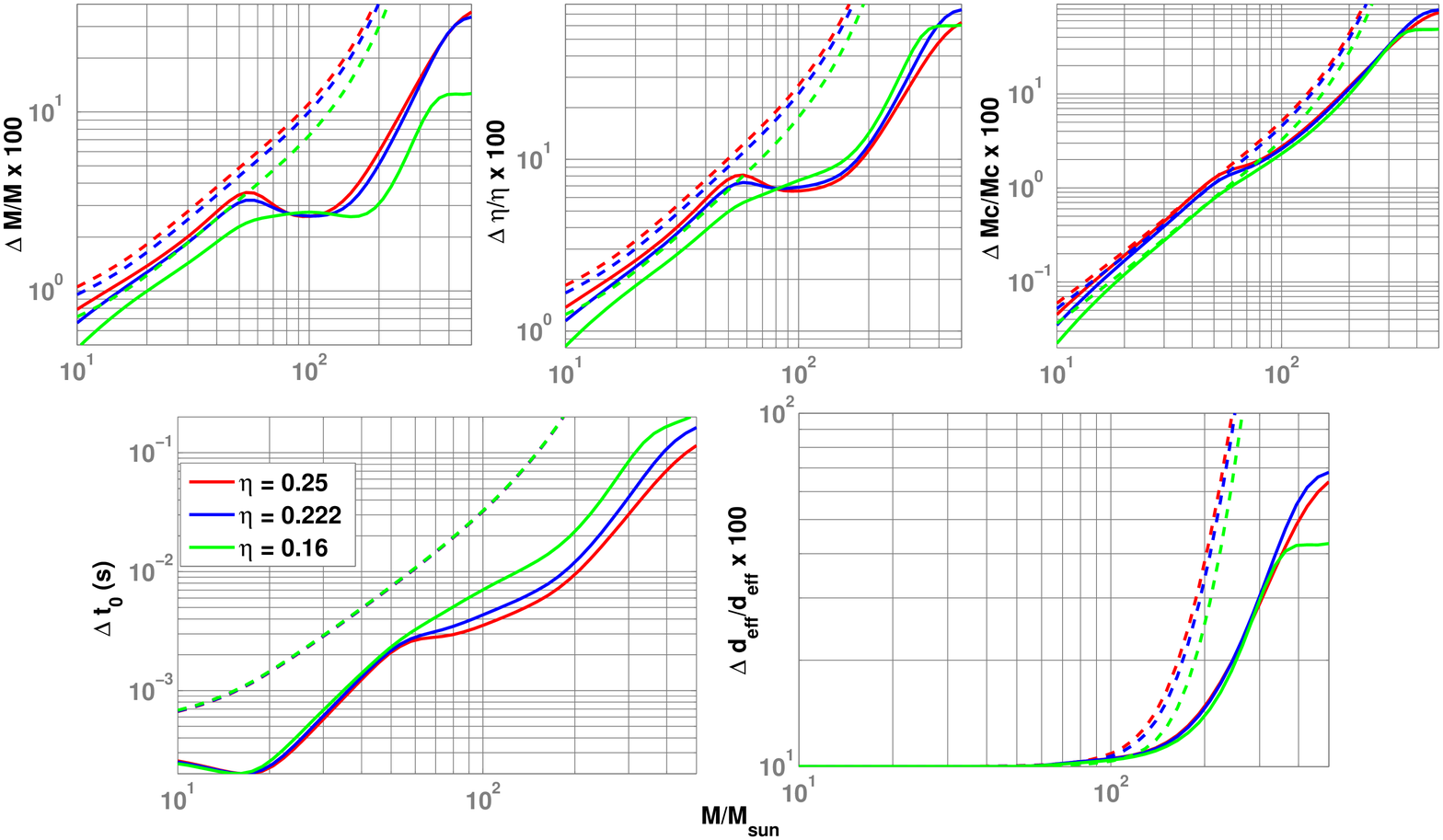}
\caption{Errors in estimating the total mass $M$ (top left), symmetric mass ratio 
$\eta$ (top middle), chirp mass $\mc$ (top right), time of arrival $t_0$ (bottom 
left) and effective distance $d_{\rm eff}$ (bottom right) in the case of Advanced 
LIGO noise spectrum, plotted against the total mass of the binary. The errors of 
$M, \eta, \mc$ and $d_{\rm eff}$ are in percentage and the errors of $t_0$ is in 
seconds. Symmetric mass-ratio $\eta$ is shown in legends. The solid lines correspond 
to a search using complete BBH templates and the dashed lines correspond to 3.5PN-accurate 
post-Newtonian templates in the SPA, truncated at the Schwarzschild ISCO. The
errors correspond to a fixed SNR of 10.}
\label{fig:FishMatErrorsFixdSNR}
\end{figure*}

\begin{figure*}[tb]
\centering
\includegraphics[width=6.1in]{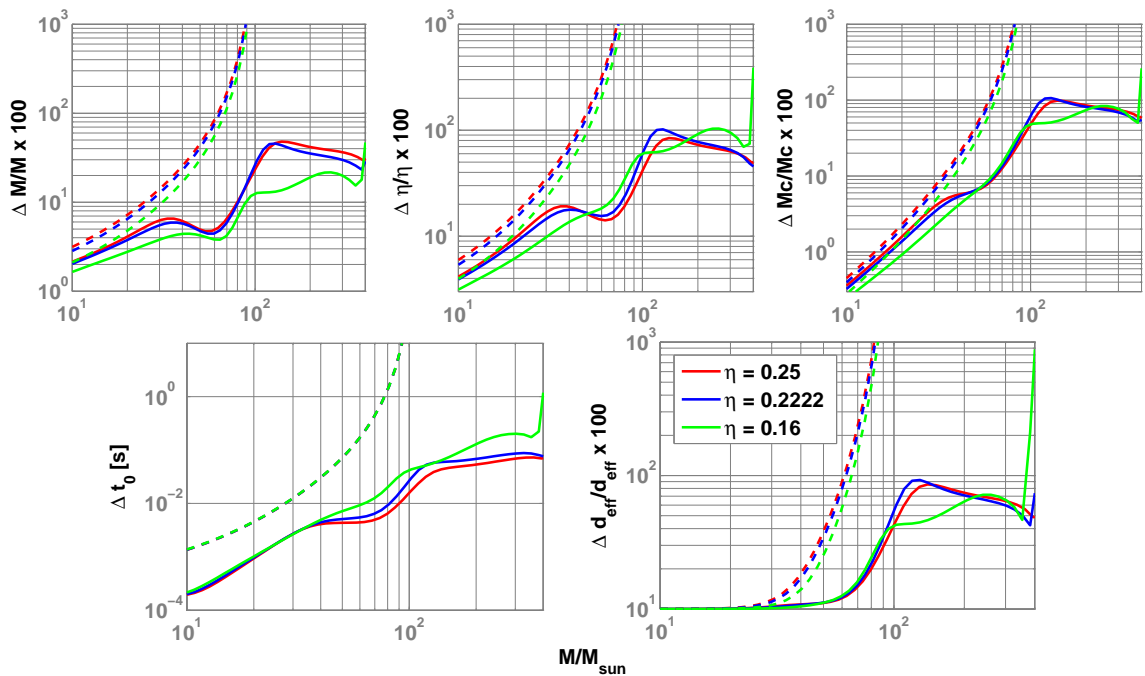}
\caption{Same as Fig.~\ref{fig:FishMatErrorsFixdSNR} except that the noise 
PSD corresponds to that of Initial LIGO.}
\label{fig:FishMatErrorsFixdSNRLIGOI}
\end{figure*}

\begin{figure*}[tb]
\centering
\includegraphics[width=6.1in]{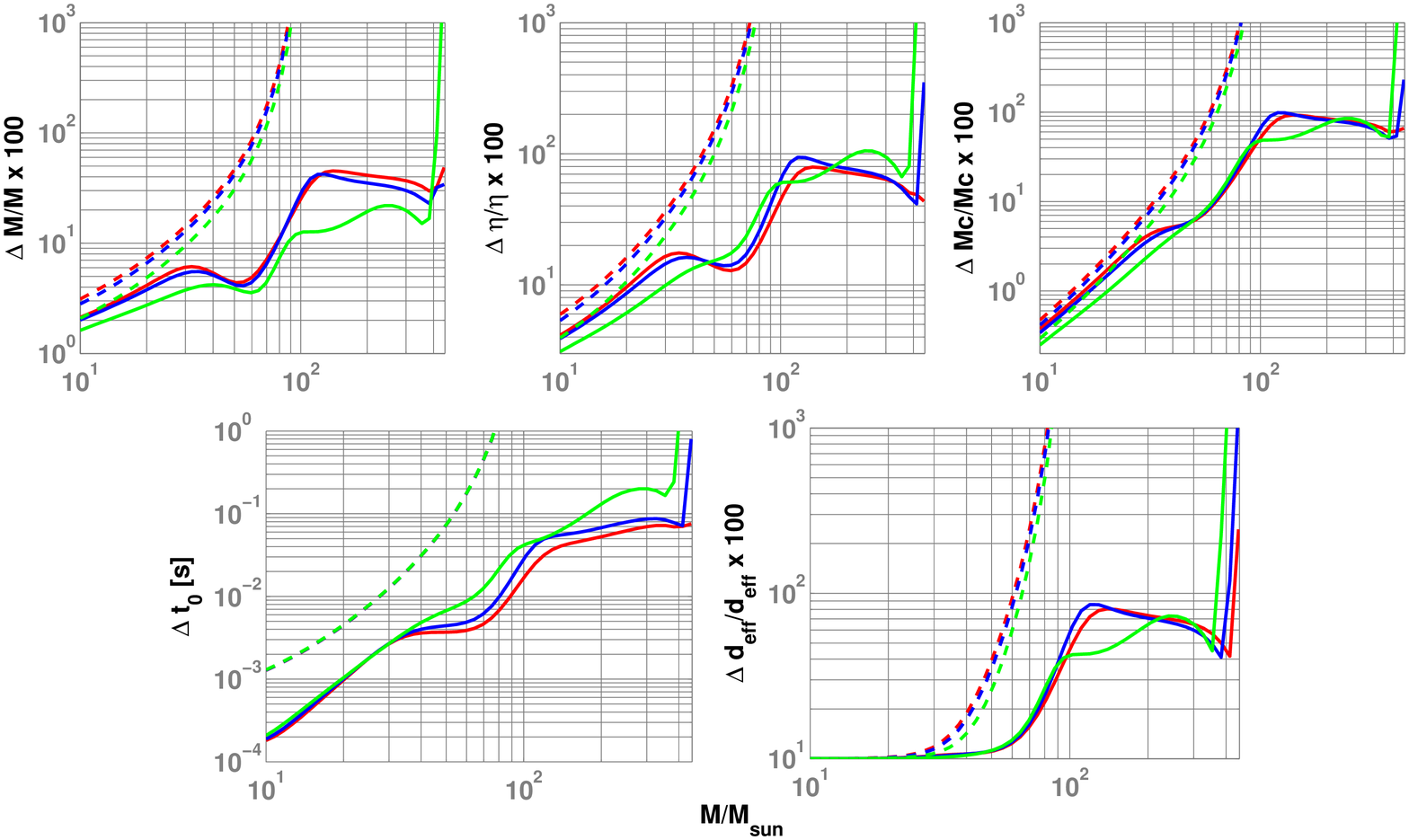}
\caption{Same as Fig.~\ref{fig:FishMatErrorsFixdSNR} except that the noise 
PSD corresponds to that of Enhanced LIGO.}
\label{fig:FishMatErrorsFixdSNREnhLIGO}
\end{figure*}

\subsection{Analytical calculation using Fisher information matrix}

In this section, we use the Fisher information-matrix formalism to estimate the 
errors in measuring the parameters of coalescing BBHs with a single GW 
interferometer. We present results for three generations of ground-based 
detectors, namely, Initial LIGO, Enhanced LIGO and Advanced LIGO.  The one-sided 
noise PSD of the Initial LIGO detector is given in terms of a dimensionless frequency 
$x=f/f_0$ by~\cite{DIS01,LAL}
\begin{widetext}
\begin{equation}
S_h\left(f(x)\right) =  9 \times 10^{-46} \left [ (4.49x)^{-56} + 0.16 x^{-4.52} + 0.52 + 0.32 x^2\right ]\,,
\end{equation}
where $f_0=150$ Hz; while the same for Enhanced LIGO reads~\cite{RanaPrivate}:
\begin{equation}
S_h\left(f(x)\right) = 1.5 \times 10^{-46} \left[ 1.33\times10^{-27} \, 
e^{-5.5\,(\ln\,x)^2}\,x^{-52.6}\,+ 0.16\,x^{-4.2} + 0.52 + 0.3\,x^{2.1} \right], 
\end{equation}
where $f_0=178$ Hz. For Advanced LIGO~\cite{LAL},
\begin{equation}
S_h\left(f(x)\right) = 10^{-49} \left [x^{-4.14} - 5 x^{-2} + 111 \Big(\frac{1 - x^2 + x^4/2}{1 
+ x^2/2}\Big)\right]\,,
\label{eq:AdLIGOPSD}
\end{equation}
where $f_0=215$ Hz, 
and, for Advanced Virgo~\cite{GLosurdo},
\begin{equation}
S_h\left(f(x)\right) = 10^{-47} \left [  2.67 \times 10^{-7} x^{-5.6} + 0.59 \, 
e^{(\ln\,x)^2\,[-3.2 - 1.08 \ln(x) 
- 0.13 (\ln\,x)^2]}\,x^{-4.1}+0.68 \, e^{-0.73 (\ln\,x)^2}\,x^{5.34} \right]\,,
\label{eq:AdVirgoPSD}
\end{equation}
\end{widetext}
where $f_0=720$ Hz. 
The calculations presented in this section were performed using the Initial LIGO,
Enhanced LIGO and Advanced LIGO noise PSDs, while the calculations presented 
in Sec.~\ref{sec:ParamEstimMultDet} consider a three-detector network consisting of 
Advanced LIGO and Advanced Virgo. 

The parameters that can be estimated through single-detector observations 
are $\{\cA(d_L), t_0, \varphi_0, M, \eta\}$. To be precise, one can measure only 
the Doppler-shifted masses, unless there are additional
experiments for determining the Doppler-shift \cite{Cutler:1994ys} and, therefore,
allow the estimation of the true masses. 
Doppler shifting can arise due to the motion of the detector relative to the 
source or the cosmological expansion. In measurements with multiple-detectors,
as discussed below, it is possible to measure the source distance and sky-position
as well. There too, the distance observed is actually the Doppler-shifted
distance. 

The Fisher matrix elements in the $\{\cA, t_0, \varphi_0, M, \eta\}$ space
are computed from the derivatives of the waveforms described by 
Eqs.(\ref{eq:phenWave}) -- (\ref{eq:ampParams}): 
\begin{widetext}
\be
\Gamma_{ab} = \left<\pa \tilde{h}(f), \pb \tilde{h}(f) \right> \simeq 4 \, 
\int_{f_{\rm low}}^{\fcut} \df ~ \frac{\pa \Aeff(f) 
\, \pb \Aeff(f) + \Aeff^2(f) \, \pa \Psieff(f) \, \pb \Psieff(f)}{S_h(f)}\,,
\ee
\end{widetext}
where the low-frequency cutoff, $f_{\rm low}$, is chosen to be 10 Hz 
for Advanced LIGO, and 40 Hz for Enhanced and Initial LIGOs. The upper-frequency cuttoff, 
$\fcut$ is given by Eq.~(\ref{eq:ampParams}). 

The rms errors in parameters $M, \eta$ and $t_0$ are computed
by inverting the Fisher matrix elements as discussed in Sec.~\ref{sec:ParamEstimTheory}. 
The error in estimating the chirp mass $\mc$ and the effective distance $\deff$
are obtained by propagating the errors in $M, \eta$ and $\cA$ in the following way: 
\ber
\left(\frac{\Delta \mc}{\mc}\right)^2 & = & \left(\frac{\Delta M}{M}\right)^2 + \frac{9}{25}
\left(\frac{\Delta \eta}{\eta} \right)^2 \nonumber \\ 
& + & \frac{6}{5} \, \corr_{M\eta} \, \frac{\Delta M}{M} \, \frac{\Delta \eta}{\eta} \\
\left(\frac{\Delta \deff}{\deff}\right)^2 & = & \frac{25}{36} \left(\frac{\Delta M}{M}\right)^2
+ \frac{1}{4} \left(\frac{\Delta \eta}{\eta}\right)^2 + \left(\frac{\Delta \cA}{\cA}\right)^2 \nonumber \\
& + & \frac{5}{6} \corr_{M\eta} \frac{\Delta M}{M} \, \frac{\Delta \eta}{\eta}
- \frac{5}{3} \corr_{M\cA} \frac{\Delta M}{M} \, \frac{\Delta \cA}{\cA} \nonumber \\
& - &  \corr_{\eta\cA} \frac{\Delta \eta}{\eta} \, \frac{\Delta \cA}{\cA}
\label{eq:ErrPropSingleDet}
\eer
where $\Delta\vt^a$ denotes the rms error in estimating $\vt^a$ obtained 
from $\Gamma_{ab}$, and $\corr_{ab}$ is the correlation coefficient between parameters $\vt^a$ and $\vt^b$.

Errors in the estimates of the parameters $M, \eta, \mc, t_0$ and $d_{\rm eff}$ in the
case of AdvLIGO detector are plotted 
against the total mass $M$ in Fig.~\ref{fig:FishMatErrorsFixdDist}. These errors are 
computed assuming that the binary is placed at an effective distance of 1 Gpc. Also 
plotted in the figures are the same error-bounds computed from the 3.5PN accurate \emph{restricted} 
PN waveforms in the stationary phase approximation (SPA), truncated at the Schwarzschild innermost stable
circular orbit (ISCO). It can be seen that, over a significant range of the total
mass, the error-bounds in the complete templates are largely better than those 
in the PN inspiral waveforms. For binaries with $M=100M_\odot$ and $\eta=0.25$, the error-bounds in 
various parameters using the complete [PN] templates are $\Delta M/M \simeq 0.34 \, [5.38] \, \%, 
~\Delta \eta/\eta \simeq 0.84 \, [12.98] \, \%,~ \Delta \mc/\mc \simeq 0.35 \, [2.47] \,\%, 
~\Delta t_0 \simeq 0.46\,[15.51]$ ms and $\Delta d_{\rm eff}/d_{\rm eff} \simeq 1.36 \, [5.24] \, \%$. 
The errors in estimating the same parameters using Initial LIGO and Enhanced LIGO detectors
are plotted in Figs.~\ref{fig:FishMatErrorsFixdDistLIGOI} and \ref{fig:FishMatErrorsFixdDistEnhLIGO}.
 
The rate of variation in the errors in different regions of 
the parameter space can be understood by studying the overlap function, which is the 
ambiguity function maximized over $t_0$ and $\varphi_0$~\cite{Sathyaprakash:1991mt}.
Figure~\ref{fig:OverlapFnContourAdvLIGO} plots the contours of the overlap between 
waveforms generated at different points in the $(M,\eta)$ space. Notice the change in the 
shape and orientation of the ambiguity ellipses, especially, as the total mass of 
the binary is varied. While, to a very good approximation, the chirp mass continues 
to remain as one of the eigen-coordinate~\cite{Sathyaprakash:1994} in the case of the low-mass (with 
$M \leq 20 M_\odot)$ binary {\em inspiral (PN)} waveforms, this is no longer true for 
the {\em complete} waveforms of higher mass systems. This is because the latter waveforms 
have more information about the component masses than just the chirp mass. The eigen-directions change 
dramatically with increasing total mass. It can be seen that the error trends reported 
in Fig.~\ref{fig:FishMatErrorsFixdDist} closely follow the shape of these ambiguity 
ellipses. This also means that while placing templates in the inspiral-merger-ring 
down searches, we will have to consider these changes in the orientation of the ambiguity 
ellipses. This will be studied in a future work. 

One common problem encountered in the estimation of errors using Fisher information
matrix is the following: In some cases (especially in the case of large number of 
parameters), the Fisher matrix becomes badly conditioned, thereby,
decreasing the fidelity of the error covariance matrix derived by inverting it.
This problem can often be obviated by intelligently choosing the parameters and by projecting out certain
dimensions in the Fisher matrix (e.g., $t_0$ and $\varphi_0$). 
We have verified our results by comparing the errors computed using the full Fisher matrix with those
computed using the projected matrix. In our calculations, they turned out to be 
the same to the extent discernible in the figures and tables presented here.

It may be noted that for a fiducial signal \emph{limited only to the inspiral phase} 
of the binary, \ie, for $f<\fmerg$, the parameter $\cA$ is uncorrelated with the 
other signal parameters, and hence one has $\Gamma_{1a}=\delta_{1a}\,\rho^2$, which 
renders the Fisher matrix in the block-diagonal form.  
However, for the complete signal, with the merger and the ringdown pieces included,
the correlation of $\cA$ with the other parameters becomes non zero, and the Fisher 
matrix is no longer block-diagonal with respect to this parameter. This implies 
that the complete waveforms provide
more information about $\cA$ and, hence, about the effective distance $d_{\rm eff}$.
 
Figures~\ref{fig:FishMatErrorsFixdSNR}, \ref{fig:FishMatErrorsFixdSNRLIGOI} and
\ref{fig:FishMatErrorsFixdSNREnhLIGO} show the error estimates corresponding to 
a fixed (single-detector) SNR of 10 in the case of Advanced LIGO, Initial LIGO 
and Enhanced LIGO noise spectra, respectively. It is interesting to note that the 
parameter estimation using the complete waveforms is still much better than that 
using only the inspiral waveform even though, in order to produce the same SNR 
using inspiral templates, the effective distance to the binary has to be often much 
smaller. The reason for this can be understood through an 
analogy with parameter estimation with multiple detectors: Since the inspiral
phase, on the one hand, and the merger-ringdown phases, on the other hand, 
occupy two contiguous and, essentially, non-overlapping frequency-bands, the 
detection of a complete signal is equivalent to a coherent detection of these 
two pieces of the waveforms by two coincident, co-aligned detectors with 
sensitivities limited to the two contiguous bands, respectively. The two
phases, however, are modulated by the two mass parameters in complementary 
ways, in the sense that the Fisher sub-matrices in the two-dimensional 
mass-space for these two fiducial detectors grow more linearly
independent of each other, the larger the total mass gets, even while the
total coherent SNR of this fiducial detector pair is held constant. This
linear independence causes the estimation of two mass parameters to 
improve. Contrastingly, since the merger-ringdown pieces add very little 
information about a system's {\em chirp-mass}, the improvement in its
accuracy arising from using complete waveforms is much less even for high
mass systems.

\begin{figure}[tb]
\centering
\includegraphics[width=3.2in]{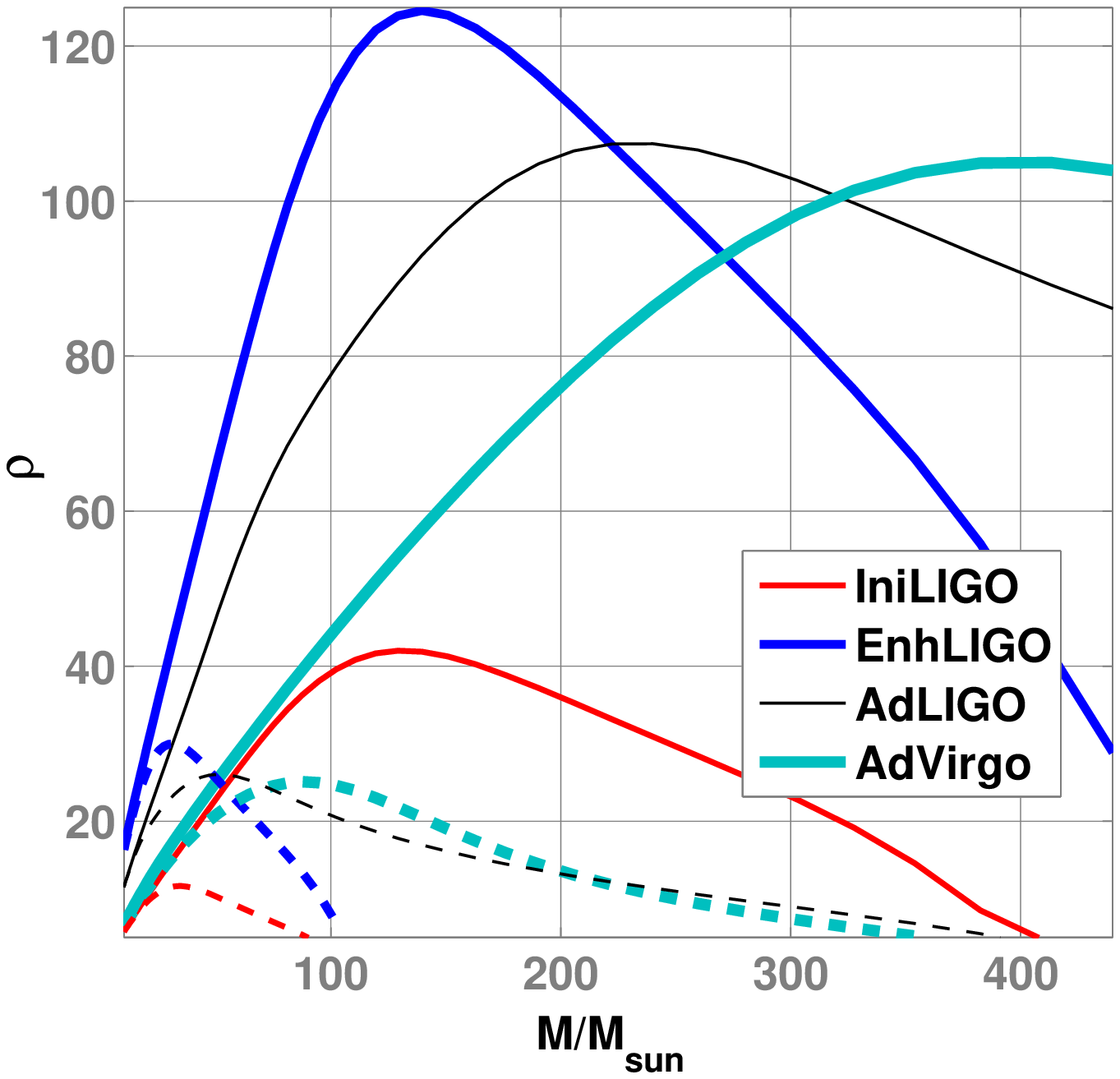}
\caption{The curves labelled ``IniLIGO'' and ``EnhLIGO'' report the SNR produced by 
binaries located at an effective distance of 100Mpc at Initial LIGO and Enhanced
LIGO, respectively, as a function of the total mass. The curves labelled ``AdLIGO'' 
and ``AdVirgo'' report the same produced by binaries located at 1Gpc at Advanced LIGO
and Advanced Virgo. The solid lines correspond to complete waveforms
and the dashed lines correspond to PN waveforms.}
\label{fig:SNRDiffDetectors}
\end{figure}

Figure~\ref{fig:SNRDiffDetectors} plots the SNR produced at different detectors 
by equal-mass binaries located at a fixed distance, as a function of the total 
mass of the binary.

\subsection{Monte-Carlo simulations}

\begin{figure}[tb]
\centering
\includegraphics[width=3.4in]{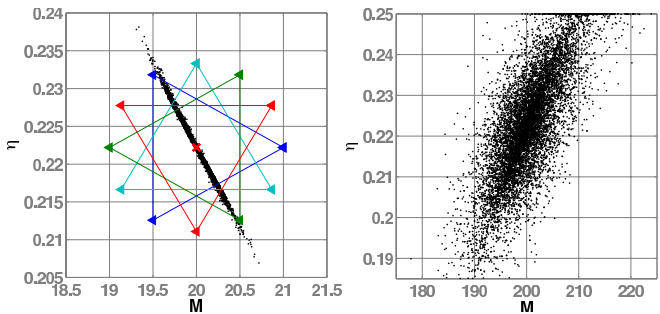}
\caption{Scatter plot of parameters estimated from $10^4$ Monte-Carlo simulations.
The horizontal axis reports the total mass and the vertical axis reports the 
symmetric mass ratio. The left panel correspond to the injection with parameters
$M=20 M_\odot$ and $\eta = 0.2222$, and the right panel correspond the injection
with parameters $M=200 M_\odot$ and $\eta = 0.2222$. The injections correspond
to an SNR of 20. Also overlaid in the left
panel is a cartoon of the four different initial simplexes chosen for the 
maximization algorithm. The true values of the parameters is marked by a cross.
Note that the eigen directions are different in the two plots.}
\label{fig:MoteCarloAdvLIGOScatterPlots}
\end{figure}

\begin{figure*}[tb]
\centering
\includegraphics[width=7.2in]{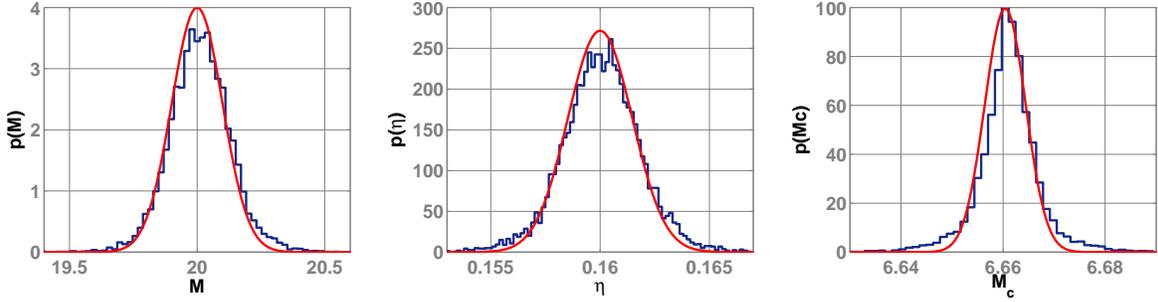}
\caption{Distribution of the estimated parameters from Monte-Carlo simulations
and expected probability distributions from Fisher matrix calculation. The true
parameters are $M=20 M_\odot, \eta = 0.16$. }
\label{fig:ErrDistMonteCarloAdvLIGO}
\end{figure*}

\begin{figure*}[tb]
\centering
\includegraphics[width=6.2in]{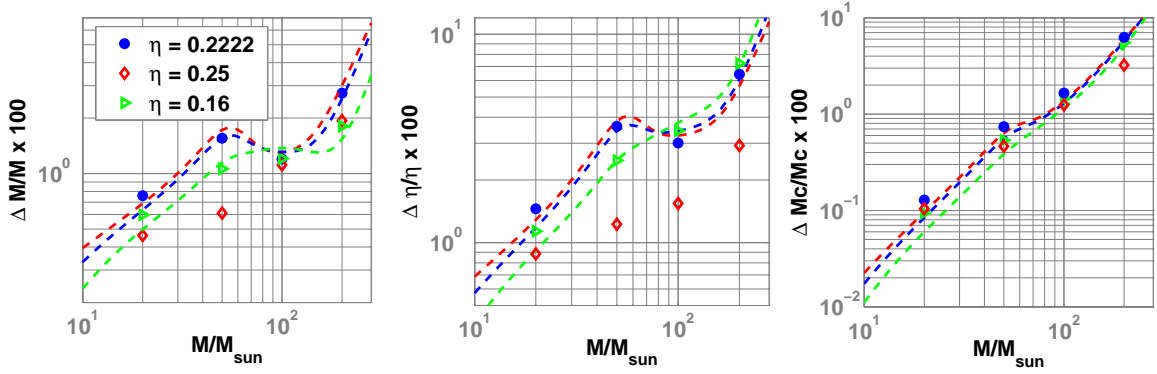}
\caption{The data points show the errors computed from Monte-Carlo simulations
in the case of Advanced LIGO noise PSD. The horizontal axis reports the total 
mass while the legends report the symmetric mass ratio. The errors are computed 
for a fixed SNR of 20. The dashed lines correspond to the same errors computed 
using Fisher matrix formalism.}
\label{fig:ErrSigmaMonteCarloVsM}
\end{figure*}

\input{ParamEstimSingDet.table}


\begin{figure*}[tb]
\centering
\includegraphics[width=6.2in]{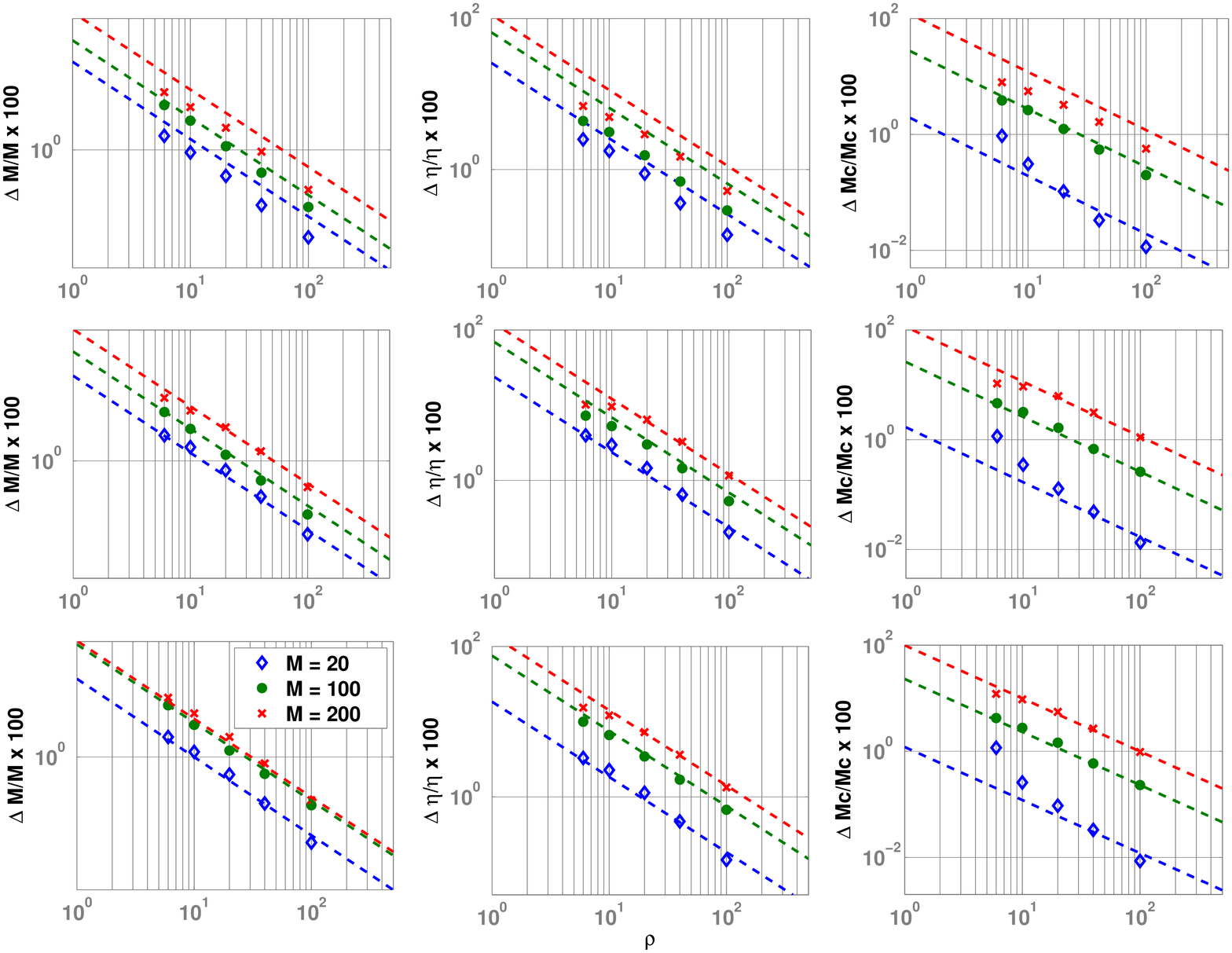}
\caption{Errors computed from Monte-Carlo simulations (crosses, dots and diamonds)
plotted against SNR. The horizontal axes report the SNR of the injections and the 
legends report the total mass in units of $M_\odot$. The top, middle and bottom 
panels correspond to mass ratios $\eta = 0.25, 0.2222$ and $0.16$, respectively. 
The error-bounds expected from the Fisher matrix calculation are indicated 
by dashed lines.}
\label{fig:ErrSigmaMonteCarloVsSNR}
\end{figure*}

The limitations of the Fisher-matrix formalism are well known~\cite{Balasubramanian:1995ff,
Balasubramanian:1995bm,Vallisneri:2007ev}. The parameter-error bounds provided 
by it are trustworthy in the limit of high SNR and for parameters on which the signal has 
linear dependence. In the case of low SNRs the error bounds computed using the Fisher 
matrix formalism can be largely different from the ``actual'' errors. Also, the Fisher 
matrix does not recognize the boundaries of the parameter space (such as the restriction 
$\eta \leq 0.25$). Neither does it account for practical restrictions such as  
the finite sampling of the data. In order to explore these limits of the Fisher formalism, 
we performed Monte-Carlo simulations, whereby maximum-likelihood detections
were made of simulated signals added to multiple statistically independent 
realizations of simulated colored, Gaussian noise. The aim of this frequentist study 
was to obtain the spread in the maximum-likelihood estimates of the parameters
and compare them with Fisher-matrix calculations.
It is worth clarifying that there is another interesting question one
can pose in the context of parameter estimation, namely, ``Given a specific
signal and a particular noise realization, what are the posterior distributions
of the parameter estimates.'' This is a question from Bayesian statistics 
that can be answered using Markov-Chain Monte-Carlo (MCMC) simulations, as
explored for inspiral-only waveforms in Refs.~\cite{Roever:2006, vanderSluys:2007st,
Raymond:2008im,vanderSluys:2008qx}. We do not answer that question here.

In this section we present results from the frequentist 
Monte-Carlo simulation studies.
These studies largely corroborate the Fisher matrix calculations in the 
parameter-space regions where the latter is expected to be trustworthy. The simulations
also allow us to compute error-bounds in the parameter-space regions 
where the Fisher matrix formalism can be unreliable (such as for $\eta \simeq 0.25$).
We caution the reader that this is not meant to be an exhaustive comparison between 
Fisher-matrix calculations and Monte-Carlo simulations. A detailed comparison of Fisher 
matrix formalism with Monte-Carlo simulations in the case of 3.5PN inspiral signals 
can be found in the recent work Ref.~\cite{CokelaerParamEstim:2008}. 

Colored Gaussian noise with one-sided PSD $S_h(f)$ is generated in the frequency 
domain. If $\hat{x}_k$ and $\hat{y}_k$ denote the real and imaginary parts of the 
discrete Fourier transform of the noise at the frequency bin $k$, these are generated by
\be
\hat{x}_k = \sqrt{S_{h_k}}\,x_k/2~,~~~\hat{y}_k = \sqrt{S_{h_k}}\,y_k/2,
\ee
where $x_k$ and $y_k$ are random variables drawn from a Gaussian distribution of 
zero mean and unit variance, and $S_{h_k}$ denotes the discrete version of $S_{h}(f)$.
Frequency domain signal described by Eq.(\ref{eq:phenWave}) is added to the noise. 
The data is filtered through a matched filter employing templates described by 
Eq.(\ref{eq:phenWave}). The likelihood is maximized over $t_0$ and $\varphi_0$
as described in Sec.~\ref{sec:PhenWave}. The maximization over the physical parameters
($M$ and $\eta$) is best performed by filtering the data using a template bank 
finely spaced in the parameter space. But, in order to attain sufficiently good 
accuracy (say, $1\%$), a large number of simulations needs to be performed. Thus, 
computing error-bounds from a good volume of the parameter space is computationally 
expensive in a template bank search. So, in this paper, the maximization over the 
physical parameters is performed with the aid of the computationally cheaper Nelder-Mead 
downhill simplex algorithm~\cite{amoeba}. 

We emphasize that this search may not be as accurate as the template bank search. One 
reason for the inaccuracy is that, in this method, we do not ``sample'' the parameter space finely 
enough, and hence the ``real maximum'' can very well be missed. This is especially the case 
when the function that we want to maximise (likelihood in this case) contains many secondary 
maxima. Indeed, it is well known that the the likelihood can have many secondary maxima 
arising due to global correlations in the parameter space. We bypass this issue by starting 
the maximisation algorithm around the ``actual'' peak of the function. Hence, the error distributions
that we obtain are only indicative of the spread of the MLM estimates around the primary maxima. 
Unlike in the case of MCMC simulations, this does not provide a complete picture of the posterior 
distribution of the parameters. Nevertheless,  this is a worthwhile tool as an independent 
verification of the Fisher matrix calculation, enabling us to ``scan''' a good volume of the 
parameter space using Monte-Carlo simulations~\footnote{In our simulations, a few hundred 
trials were sufficient for the Nelder-Mead's algorithm to converge to the fiducial maximum. By 
contrast, a template bank search requires tens of thousands of templates, in general.}. 

Nelder-Mead's algorithm is a multidimensional minimisation/maximisation algorithm. In order 
to maximize the required function, we need to specify an initial ``simplex'' of $n+1$
dimensions where $n$ is the dimensionality of the parameter space. Since the dimensionality
of our parameter space is 2, the simplex in our case is a triangle. It is important
for the good convergence of the maximization that the initial simplex ``catch'' the 
orientation of the ambiguity ellipses in our parameter space, which often depends 
strongly on the parameters themselves. Thus, we start the maximization by specifying 
four different initial simplexes, whose vertices have equal (coordinate) distance from the 
``true'' value of the parameters. The four triangles are oriented in different directions in 
the parameter space. We choose the parameters corresponding to the best among the 
maximized likelihoods as the parameters of the injection. Figure~\ref{fig:MoteCarloAdvLIGOScatterPlots} 
shows a scatter plot of the parameters estimated from $10^4$ simulations. Also overlaid 
in the left plot is a cartoon of the initial simplexes chosen. The reader may note the difference 
in the eigen-directions in the two plots. 

We found that the following points need to be taken care of while performing this kind of 
simulations: (i) Since the frequency-domain templates are abruptly cut off at the 
frequency $\fcut$, we need to make sure that the edges arising from this do not corrupt 
our numerical calculations. This means that, for high mass systems ($M > 200 M_\odot$) 
we cannot perform the simulations with very high SNR ($\rho > 100$), because the 
cutoff frequency is at the ``sweet spot'' of the detector. (ii) Sufficiently small
tolerance level for the maximization algorithm in order to ensure that the ``true'' 
maximum is never missed. (iii) Orthonormality of the search templates, as emphasized
by Ref.~\cite{Balasubramanian:1995bm}. 

The frequency distributions of the estimated parameters $M, \eta$ and $\mc$ 
are shown in Fig.~\ref{fig:ErrDistMonteCarloAdvLIGO}. The injection corresponds to the parameter 
values $M=20 M_\odot$ and $\eta = 0.16$ and an SNR of 20. Also plotted in the figures are 
the expected distributions computed using the Fisher matrix 
formalism. All the results are computed using the AdvLIGO noise PSD. It can be 
seen that the two calculations agree very well. Figure~\ref{fig:ErrSigmaMonteCarloVsM} shows 
the errors computed using the Monte-Carlo simulations plotted against the total mass of 
the binary for three different values of $\eta$. The simulations are performed
with an SNR of 20. Also shown are the error-bounds computed using the Fisher matrix 
formalism. In the case of mass ratios $\eta = 0.2222$ and $\eta = 0.16$, the simulations
agree well with the Fisher matrix calculations. But the simulations disagree with the 
Fisher calculations for the case of $\eta = 0.25$. This is expected because the Fisher 
matrix does not recognize the physical restriction that $\eta$ can only take values 
less than, or equal to 0.25. The Fisher matrix calculation assumes that the errors in 
estimating the parameters are Gaussian distributions centered around $\eta = 0.25$, while 
the Monte-Carlo simulations enforce the restriction $\eta \leq 0.25$. As a result the 
error bounds estimated by the Monte-Carlo simulations will be less
than that estimated by the Fisher matrix. 

Fisher matrix calculations assume that the errors decrease inversely proportional to the 
SNR. But this approximation is not valid at low SNRs. So we have performed Monte-Carlo
simulations with various SNRs in order to study the SNR dependence of the errors. 
Figure~\ref{fig:ErrSigmaMonteCarloVsSNR} plots the errors estimated from the simulations
against the SNR of the injections. The top, middle and bottom panels in the figure 
correspond to mass ratios $\eta = 0.25, 0.2222$ and $0.16$, respectively. The different
markers correspond to the Monte-Carlo simulations and the dashed lines correspond to the 
Fisher matrix calculations. It can be seen that, barring the case of $\eta = 0.25$, 
the simulations agree very well with the Fisher calculations in the limit of high SNRs 
($\rho > 10$). Because of the $\eta$-boundary effects, the errors computed from the 
$\eta=0.25$ simulations are less than those computed from the Fisher calculations. For small SNRs
($\rho \leq 10$), the simulation errors start to deviate from the Fisher calculations. 
There are two reasons for this: (i) at low SNRs, as observed by many others (see, for e.g., 
Ref.~\cite{Balasubramanian:1995bm}) the Fisher matrix largely underestimates the errors.
This is the dominating effect in the case of $M=20M_\odot$ binaries at low SNRs in 
Fig.~\ref{fig:ErrSigmaMonteCarloVsSNR}.
(ii) at low SNRs, since the size of the ambiguity ellipses are increased, they are cut by
the $\eta = 0.25$ boundary, which is neglected by the Fisher calculations. Hence the 
Fisher matrix over estimate the errors. This is the dominating effect in the case of 
$M=200M_\odot$ binaries at low SNRs. It is the interplay between these two competing 
effects that causes the discrepancy between the simulations and Fisher calculations. 
In summary, the results from the Monte-Carlo simulations, albeit the limitations of the 
maximization algorithm used, should be more reliable than the Fisher calculations. 

Table~\ref{tab:ParamEstimErrorsAdvLIGO} tabulates the errors in the case of Advanced
LIGO noise PSD, computed using both Fisher matrix and Monte-Carlo simulations.

\section{Parameter estimation: Multi-detector search}
\label{sec:ParamEstimMultDet}

\begin{figure}[tb]
\includegraphics[height=6.8cm]{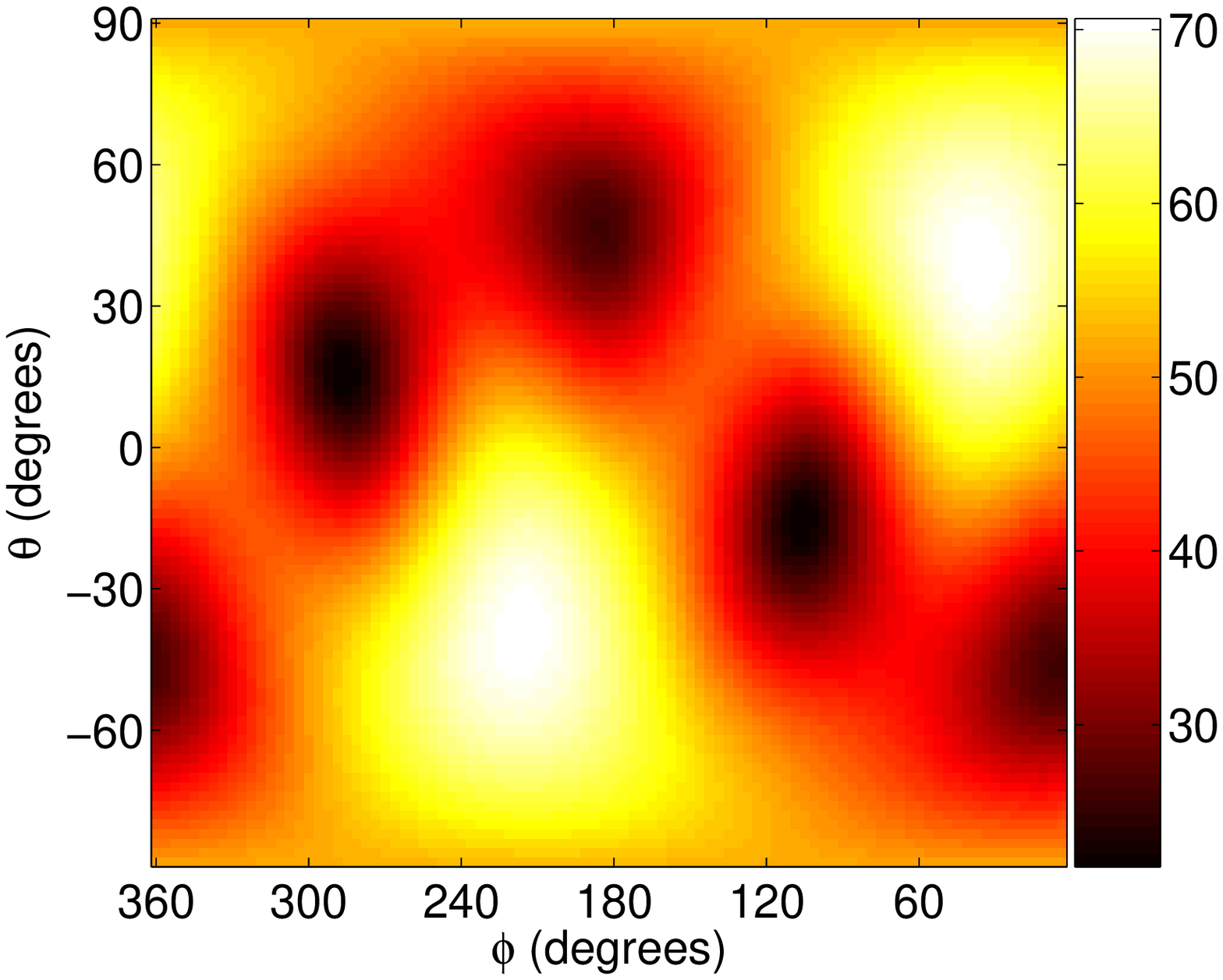}\\
\caption{The network SNR of a signal, corresponding to the 
complete waveform, from an equal-mass binary with $M=100M_\odot$ 
located at $d_L$ = 1Gpc, plotted as a function of its sky-position.
The network here is the three detector AdvLIGO-AdvVirgo network, such that
the two 4km-arm-length LIGO detectors in Hanford and Livingston have
AdvLIGO noise PSDs and the Virgo detector in Cascina has AdvVirgo noise PSD.
Above, $\theta$ and $\phi$ are the polar and azimuthal angles specifying the location 
of the source in the sky in the geographic coordinate system.}
\label{SNRskymapsM100Complete}
\end{figure}


\begin{figure*}[tb]
\includegraphics[height=6.6cm]{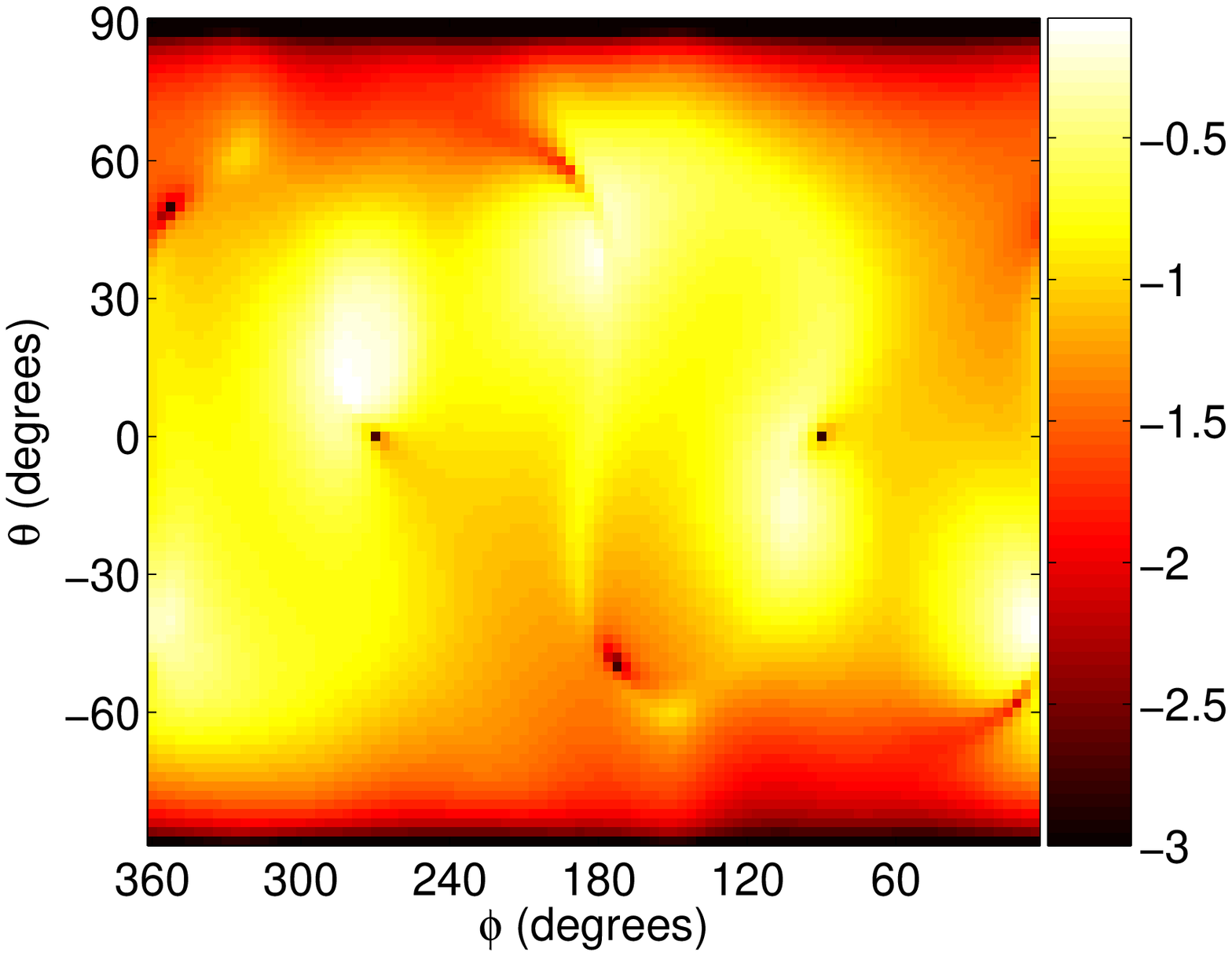}
\includegraphics[height=6.6cm]{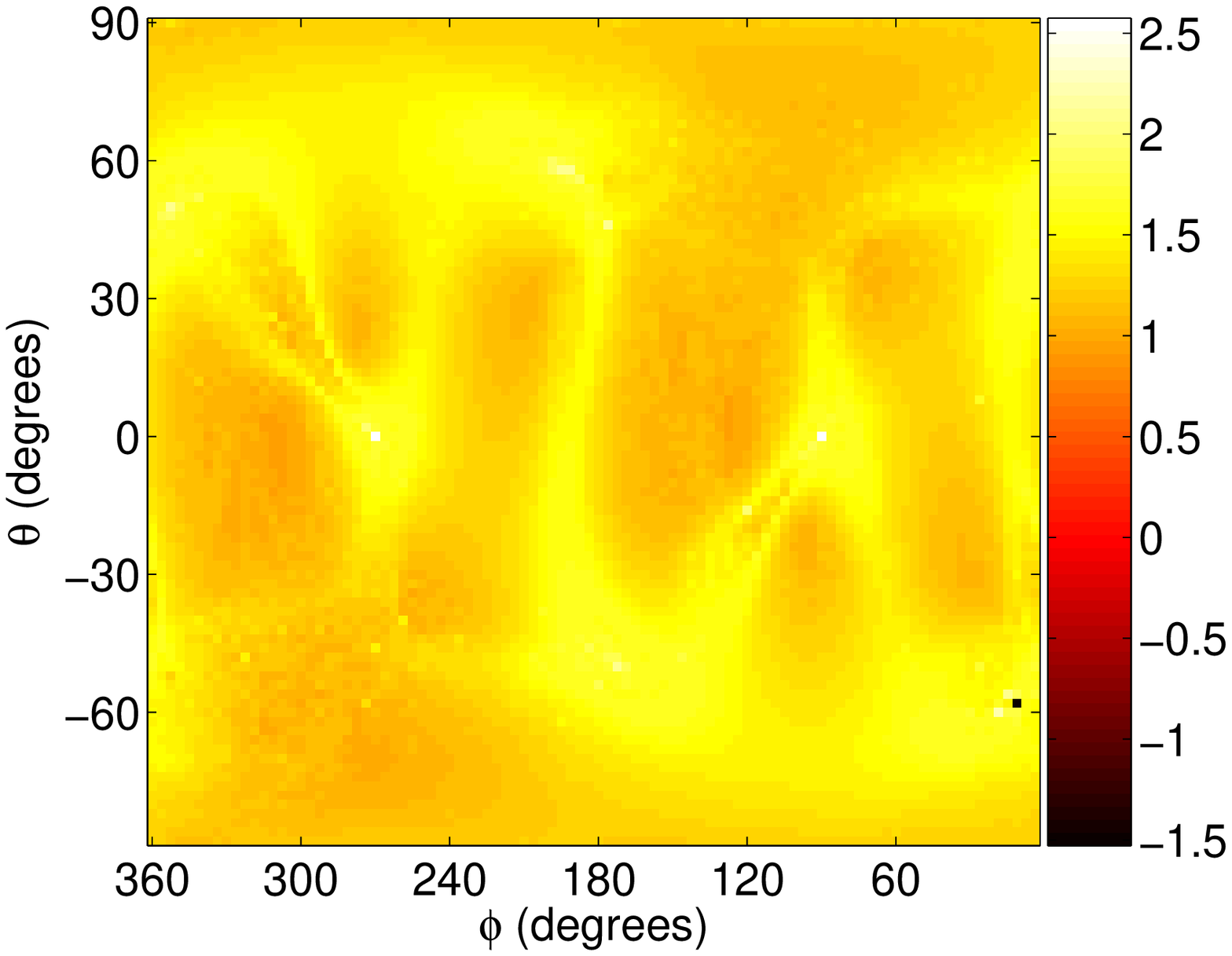}
\caption{The left plot shows the sky-position error $\log_{10}$[$\Delta\Omega$ 
(in square-degrees)] and the right plot shows the fractional error in 
the luminosity distance $\log_{10}$[$\Delta d_L/d_L$ (in \%)] as functions
of the sky-position of a BBH source. The source studied here is the
same equal-mass binary considered in Fig.~\ref{SNRskymapsM100Complete},
and, $\theta$ and $\phi$ are the polar and azimuthal angles specifying the location 
of the source in the sky in the geographic coordinate system.
Note how the effect of the varying network sensitivity, as seen in the SNR plot 
in Fig.~\ref{SNRskymapsM100Complete}, is imprinted in the two error plots. 
Additionally, the error plots display a full ``sine-wave'' pattern, which 
comprises a set of sky-positions for which the geometric 
independence of the LIGO-Virgo detectors is the weakest. Extraction of the 
signal's polarization is affected the most at these locations. That in turn 
hurts the distance measurement accuracy. The same locations do not necessarily 
hurt the determination of the sky-position, which is mostly driven by the 
measurement accuracy of the times-of-arrival of the signal at the three sites.
}
\label{ErrskymapsM100Complete}
\end{figure*}


\begin{figure}[htb]
\centering
\includegraphics[width=3.4in]{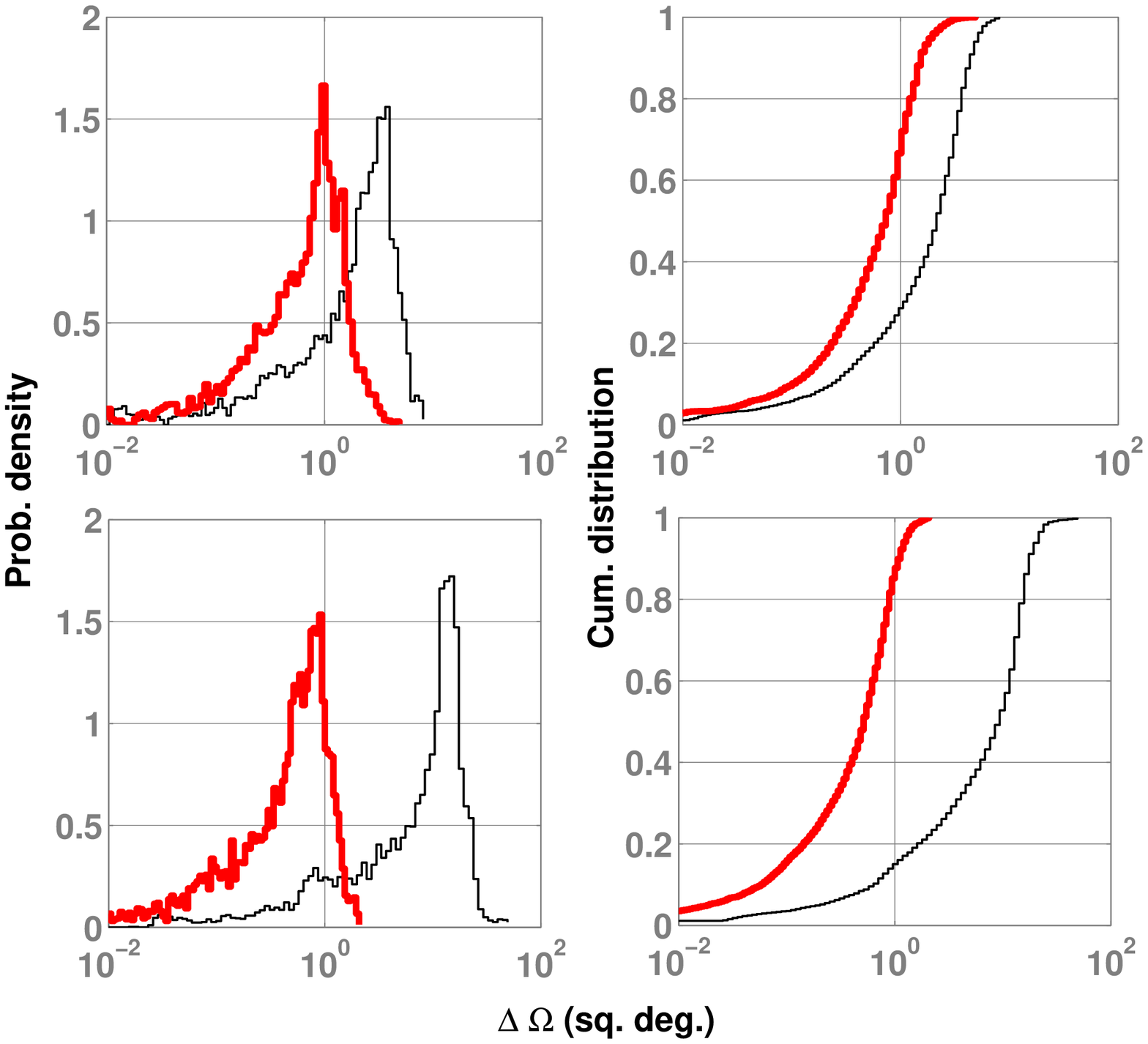}
\caption{All-sky distribution of errors in estimating the solid angle $\Omega$ in 
the case of AdvLIGO-AdvVirgo network. The left plots show the probability density 
and the right plots show the cumulative distribution. The top panels correspond to 
an equal-mass binary with $M=20 M_\odot$, and the bottom panels to one with $M=100 M_\odot$. 
In each plot the thick (red) traces correspond to the errors estimated using the complete 
waveforms while the thin (black) traces correspond to those estimated using restricted 
3.5PN waveforms. All the errors are computed for a network SNR of 10 for the respective 
waveforms.}
\label{fig:dOmegaDistributionsAdvLIGOAdvVirgoFixdSNR10}
\end{figure}

\begin{figure}[htb]
\centering
\includegraphics[width=3.4in,]{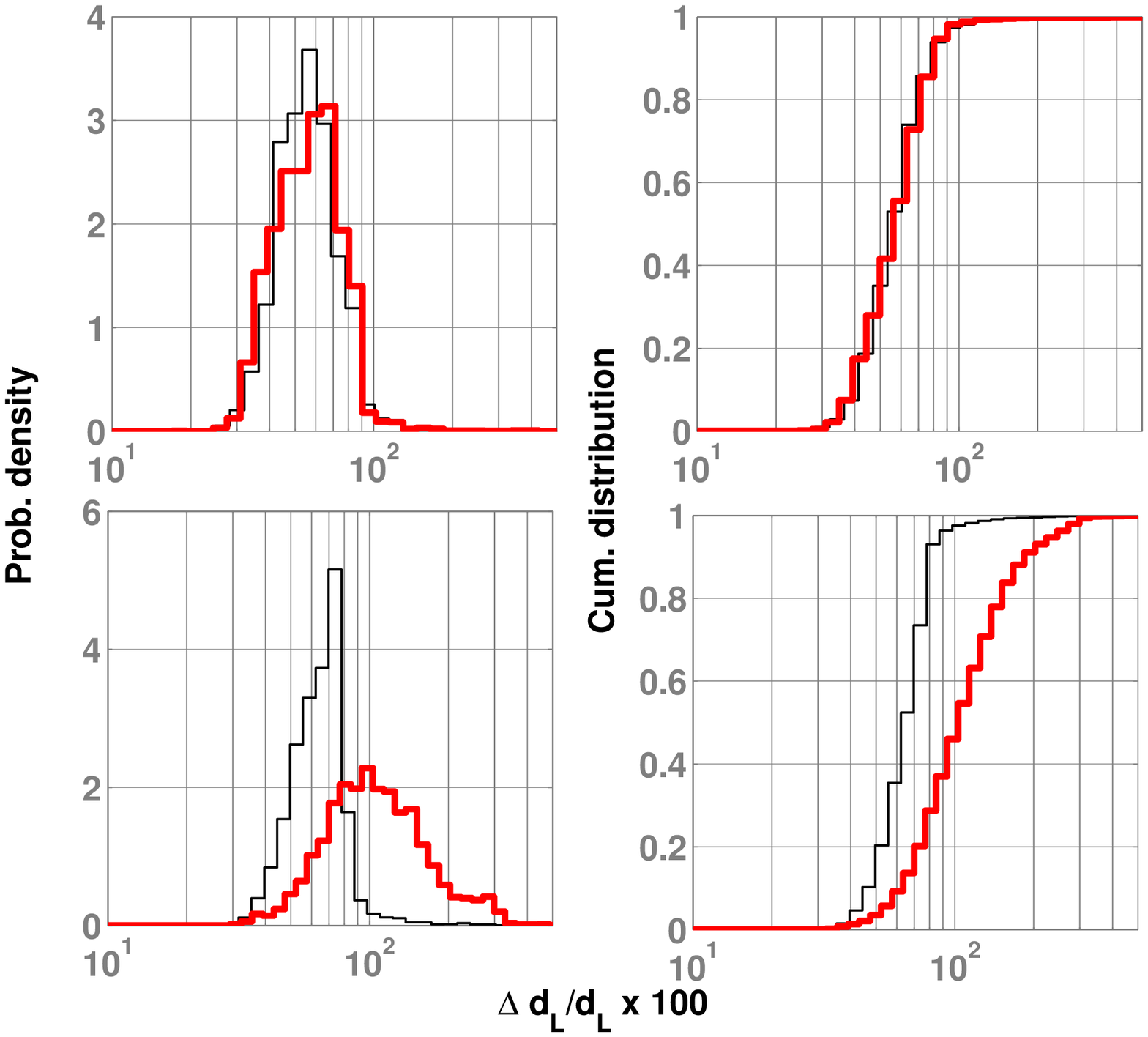}
\caption{All-sky distribution of errors in estimating $d_L$ in the case of AdvLIGO-AdvVirgo 
network. The left plots show the probability density and the right plots show the 
cumulative distribution. The top panels correspond to an equal-mass binary with 
$M=20 M_\odot$, and the bottom panels to one with $M=100 M_\odot$. In each plot the 
thick (red) traces correspond to the errors estimated using the complete waveforms
while the thin (black) traces correspond to those estimated using restricted 3.5PN 
waveforms. All the errors are computed for a network SNR of 10 for the respective 
waveforms.}
\label{fig:dLDistributionsAdvLIGOAdvVirgoFixdSNR10}
\end{figure}

\begin{figure}[htb]
\centering
\includegraphics[width=3.4in]{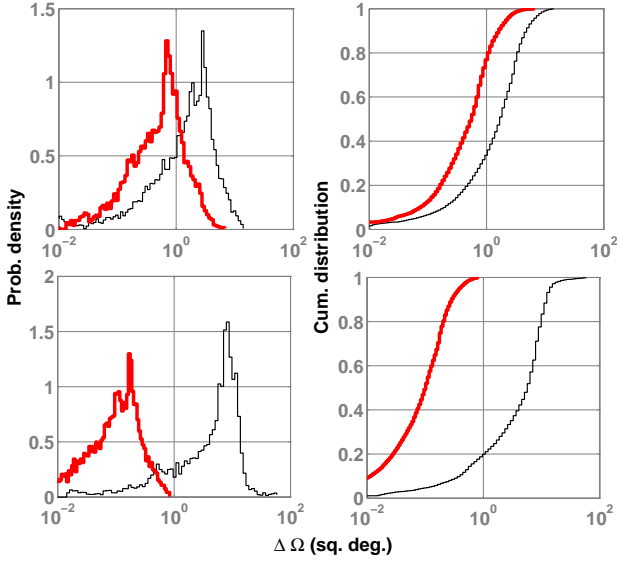}
\caption{Same as Fig.~\ref{fig:dOmegaDistributionsAdvLIGOAdvVirgoFixdSNR10} except
that the binary is now placed at a fixed luminosity distance of 1Gpc. Notice the strong 
similarity between the plots in the top panel above and those in the top panel of 
Fig.~\ref{fig:dOmegaDistributionsAdvLIGOAdvVirgoFixdSNR10}. This is because in the  
plots of the top panel above the average SNR is relatively close to 10.
The plots in the bottom rows of the two figures are more disparate: The average SNR above 
is several [few] times better than the fixed SNR in Fig.~\ref{fig:dOmegaDistributionsAdvLIGOAdvVirgoFixdSNR10} 
for the complete [inspiral-only] waveforms.}
\label{fig:dOmegaDistributionsAdvLIGOAdvVirgoFixdD1Gpc}
\end{figure}

\begin{figure}[htb]
\centering
\includegraphics[width=3.4in]{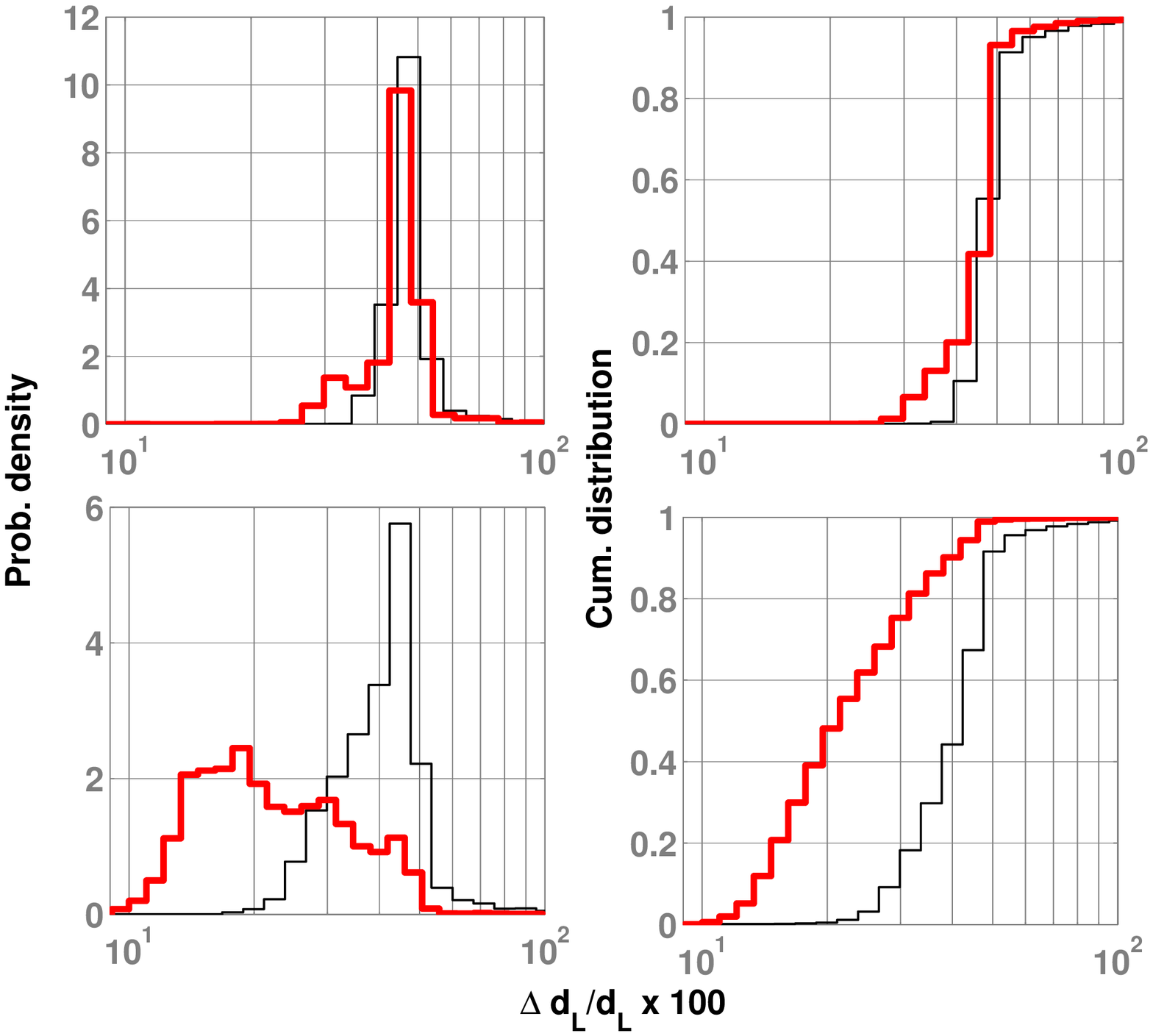}
\caption{Same as Fig.~\ref{fig:dLDistributionsAdvLIGOAdvVirgoFixdSNR10} except
that the binary is placed at a fixed luminosity distance of 1Gpc. By comparing the 
above figure with Fig.~\ref{fig:dLDistributionsAdvLIGOAdvVirgoFixdSNR10}, it is manifest 
that nearly all the improvement in the luminosity-distance measurement accuracy, when 
including the post-inspiral phases, arises due to the increased SNR. }
\label{fig:dLDistributionsAdvLIGOAdvVirgoFixdD1Gpc}
\end{figure}

With a sufficiently large number of geometrically independent 
and well-separated interferometric detectors it is possible to measure 
all nine of the BBH parameters of an adequately strong source 
\cite{Bose:1999pj,Pai:2000zt}. To assess how accurately such a 
measurement can be made with the AdvLIGO-AdvVirgo network, one can begin by 
computing the Fisher matrix in the nine-dimensional parameter space, 
and then invert it to obtain the error variance-covariance matrix.
We take the network to comprise three interferometers, with one 
each at Hanford (WA), USA, Livingston (LA) USA, and Cascina, Italy.
The LIGO detectors in Hanford and Livingston are assumed to 
be having the AdvLIGO noise PSDs given in Eq.(\ref{eq:AdLIGOPSD}) and the 
Virgo detector in Cascina is assumed to be having the AdvVirgo noise PSD given in 
Eq.(\ref{eq:AdVirgoPSD}).

When interpreting the astrophysical implications of these parameter errors, 
it is important to remember that it is only when the signal is linear in 
the parameters or the SNR is large that the maximum-likelihood estimator 
is unbiased and the error deduced from the Fisher matrix achieves the 
Cram\'{e}r-Rao bound \cite{Helstrom}. To aid this conformity, we map 
four of the six {\em extrinsic} signal parameters (i.e., parameters that depend
on the observers location in time and space), viz.,
$(\cA, \psi, \iota, \varphi_0)$, into new parameters, $a^k$, with 
$k=$1,...,4, such that the signal in Eq. (\ref{eq:hOfTDomMode}) at any 
given detector has a {\em linear} dependence on them:
\be
\label{eq:hOfTDomModeInAs}
h(t) = \sum_{k=1}^4 a^k {\sf h}_k(t) \,,
\ee
where the ${\sf h}_k(t)$'s are completely independent of those four extrinsic 
parameters. (The two remaining extrinsic parameters are the sky-position 
angles.) To deduce their dependencies as well as the forms of the $a^k$'s
we begin by noting that the antenna-pattern functions can be treated as the 
components of a vector that are related to two sky-position dependent functions, 
$u(\theta,\phi)$ and $v(\theta,\phi)$~\cite{Jaranowski:1996hs,
Pai:2000zt}, through a two-dimensional rotation by $2\psi$:
\be
\left(\begin{array}{c} F_+ \\ F_\times \end{array}\right) 
= \left(\begin{array}{cc} \cos 2\psi & \sin 2\psi \\ -\sin 2\psi & \cos 2\psi \end{array}\right) 
\left(\begin{array}{c} u \\ v \end{array}\right) \,.
\ee
With this well-known observation, one finds
\bea
{\sf h}_1(t) &\propto& u(\theta,\phi)\cos[\varphi(t)] \,, \noQ
{\sf h}_2(t) &\propto& v(\theta,\phi)\cos[\varphi(t)] \,, \noQ
{\sf h}_3(t) &\propto& u(\theta,\phi)\sin[\varphi(t)] \,, \noQ
{\sf h}_4(t) &\propto& v(\theta,\phi)\sin[\varphi(t)] \,,
\eea
where the proportionality factor is a dimensionless (mass-dependent) function 
of time. 

The new parameters are themselves defined as
\be\label{eq:DefAs}
{\cal M}_a \equiv \left(\begin{array}{cc} a^1 & a^3 \\ a^2 & a^4 \end{array}\right)
= \frac{y(\iota)}{d_L} ~{\cal O}_{\varphi_0}\cdot {\cal I} \cdot {\cal O}_{2\psi} \,,
\ee
where $y(\iota)\equiv \left[ \left(1 + \cos^2\iota \right)^2 + 4\cos^2\iota\right]^{1/2}$, ${\cal O}_\alpha$ 
is the two-dimensional orthonormal rotation matrix for angle $\alpha$ and
\be
{\cal I} \equiv \left(\begin{array}{cc} 
\left(1 + \cos^2\iota \right) / y(\iota) & 0 \\
0 & 2\cos\iota/ y(\iota) \end{array}\right) \,.
\ee 
The Fisher matrix is then computed on the space
$(M,\eta,\theta,\phi, t_0, a^1, a^2, a^3, a^4)$.
The errors in the $a^k$'s are obtained by inverting that matrix.
By using error-propagation equations obtained from Eq. (\ref{eq:DefAs}), we
are able to deduce error estimates for all four extrinsic parameters. 

In this paper, however, we present the error estimates for, perhaps, the most 
astrophysically interesting of those, namely, the luminosity distance. To obtain
it, first notice that 
\be
{\rm tr} \left( {\cal M}_a^{\bf T} ~{\cal M}_a \right) = \| {\bf a} \|^2 = \frac{y^2(\iota)}{d_L^2} \,,
\ee
where ${\rm tr}$ is the trace, and 
$\| {\bf a} \|^2 \equiv \sum_{k=1}^4 \left( a^k\right)^2$. This yields
\ber\label{eq:ErrDist}
\frac{\rmd \left(d_L\right)}{d_L} = \frac{\rmd y}{y} - \frac{\rmd \| {\bf a} \|}{\| {\bf a} \|} \,,
\eer
which can then be used to deduce the rms error, $\Delta d_L/d_L$
by accounting for the covariance between $y$ and $a^k$. Finally, we choose a flat 
prior in $y(\iota)$, such that whenever its estimate is negative or greater than 
its maximum possible value (of four) the prior is set to zero. The distance errors
plotted below are for such a prior.

The error variance-covariance matrix described above can also be used to derive
the error estimates for the other astrophysically interesting quantity, namely, the
sky-position. Here again, to further keep our assessment robust, we first
reduce the dimensionality of the Fisher matrix to five by projecting out
the four above-mentioned extrinsic parameters.
This helps in lowering the condition number
of the Fisher matrix across the parameter space.
We do so by taking a cue from Refs. \cite{Bose:1999pj,Pai:2000zt}, 
where it was shown that the network likelihood ratio of compact binary
inspiral signals can be maximized analytically over those four extrinsic 
parameters. Moreover, just as for the signal in a
single detector, it is possible to speed up the search in $t_0$ by using the 
FFT~\cite{Pai:2000zt}. Thus, the only parameters that
need to be searched numerically through the help of a template bank~\cite{Owen:1995tm}
are the following four parameters: $(M,\eta,\theta,\phi)$.

The resulting Fisher matrix is well-behaved everywhere in the five-dimensional sub-space 
except on a set of points of measure zero, where the detectors in the network cease to be 
geometrically independent. Its inverse yields the error estimates for the two mass 
parameters and the sky-position. A sky-map of the network SNR is presented in 
Fig.~\ref{SNRskymapsM100Complete} while the sky-maps of the errors in the source 
luminosity-distance and the sky-position are given in Fig.~\ref{ErrskymapsM100Complete} 
for equal-mass BBH sources with M=100$M_\odot$ and located at $d_L=1$Gpc.

Figure~\ref{fig:dOmegaDistributionsAdvLIGOAdvVirgoFixdSNR10} shows the all-sky distribution of 
the errors in estimating the solid angle $\Omega$. The left plots show the probability density and the 
right plots show the cumulative distribution. We assume that the sources are distributed uniformly 
across the sky. Top panels correspond to a binary with 
$M=20 M_\odot$ and $\eta=0.25$, and the bottom panels to a binary with $M=100 M_\odot$ 
and $\eta=0.25$. In each plot the thick (red) traces correspond to the errors estimated 
using the complete waveforms while the thin (black) traces correspond to those estimated 
using restricted 3.5PN waveforms in the SPA truncated at Schwarzschild ISCO. All the errors are 
computed for a network SNR of 10 for the respective waveforms. The error-estimates are obtained by 
averaging over the angles $(\psi,\iota)$. These plots show that in the case of an
$M=20 M_\odot$ and $\eta=0.25$ binary, assuming that the sources are distributed uniformly across 
the sky, the sky-position of 70\%\,[10\%] of the sources can be estimated with an accuracy 
better than 1\,[0.1] square degree. Using PN templates, the sky-location of only 29\%\,[6\%] of 
the sources can be estimated with an accuracy better than 1\,[0.1] square degree. 
For the $M=100 M_\odot$ binary, the sky-position of 90\%\,[18\%] of the sources can be estimated
with an accuracy of  1\,[0.1] square degree using complete waveforms, while only 15\%\,[4\%]
of the sources can be resolved with the same accuracy using inspiral waveforms. It should be 
noted that in that figure we have normalized the errors for SNR fixed to 10. For real systems additional 
improvement might be seen from the use of the complete waveforms provided their inclusion of 
merger and ringdown phases actually improves the SNR of those signals. This is indeed the case 
for high-mass systems ($M>20 M_\odot$). For an equal-mass binary with $M=20\,[100]\,M_\odot$, 
the improvement in the SNR by the inclusion of merger and ringdown is $9\%\,[300\%]$, in stationary, 
Gaussian noise. (See the discussion of Fig.~\ref{fig:dOmegaDistributionsAdvLIGOAdvVirgoFixdD1Gpc} below.)

Figure~\ref{fig:dLDistributionsAdvLIGOAdvVirgoFixdSNR10} shows the distribution of the errors 
in estimating the luminosity distance $d_L$ to two different types of equal-mass binary systems, 
both producing a network SNR of 10 in the AdvLIGO-AdvVirgo network. The top panels correspond to a 
binary with $M=20 M_\odot$ and the bottom panel to a binary with $M=100M_\odot$. As in 
Fig.~\ref{fig:dOmegaDistributionsAdvLIGOAdvVirgoFixdSNR10}, the thick (red) traces correspond 
the complete waveforms while the thin (black) traces correspond to the PN waveforms. 
These plots suggest that for an SNR of 10 the luminosity distance to around $10\%[50\%]$ of the 
sources can be estimated with an accuracy of better than $38\%[53\%]$ in the case of low-mass 
systems. They also reveal that, for a fixed value of the network SNR, the error estimates using inspiral 
and complete waveforms are almost identical. This is not surprising because for low-mass systems 
the signal is dominated by the inspiral phase. In the case of high-mass systems with an SNR of 10, 
the luminosity distance to around $10\%[50\%]$ of the sources can be estimated with an accuracy of 
$60\%[100\%]$. These errors are worse than those for the PN waveform~\footnote{Note that, in order
to get the same SNR in the case of PN waveforms, the binary must be placed at a much closer distance.} 
primarily because the covariances between the initial phase and $(\psi,\iota)$ are stronger in the 
case of complete waveforms. This property of the complete waveforms mitigates the 
estimation accuracy of $\iota$, which, in turn, affects the estimation of $d_L$.

Figures~\ref{fig:dOmegaDistributionsAdvLIGOAdvVirgoFixdD1Gpc} and \ref{fig:dLDistributionsAdvLIGOAdvVirgoFixdD1Gpc}
show the errors in estimating $\Omega$ and $d_L$ in the case of 
binaries distributed uniformly across the sky but located at a luminosity distance of 1Gpc. 
These errors also are averaged over $\psi$ and $\iota$. These plots show that
in the case of an equal-mass binary with $M=20M_\odot$
the sky-position of around $10\%[50\%]$ of the sources can be estimated with 
a resolution of $0.07[0.5]$ square degree or better. In the case of a $M=100M_\odot$ binary, 10\%[50\%] of 
the sources can be estimated with a resolution of 0.01[0.1] square degrees. These plots in 
Fig.~\ref{fig:dOmegaDistributionsAdvLIGOAdvVirgoFixdD1Gpc} also show that the coherent addition of 
the merger and ringdown phases brings about remarkable improvement (i.e., by several times for 
most sky-positions) in the estimation of $\Omega$. 

Figure~\ref{fig:dLDistributionsAdvLIGOAdvVirgoFixdD1Gpc} shows that the luminosity distance of $10\%[50\%]$ 
of the $M=20M_\odot$ BBH sources can be estimated with $32\%[47\%]$ accuracy or better and that of 
10\%[50\%] of the $M=100M_\odot$ binaries can be estimated with an accuracy of 13\%[20\%] or better. 
While comparing Figs.~\ref{fig:dLDistributionsAdvLIGOAdvVirgoFixdSNR10} and 
\ref{fig:dLDistributionsAdvLIGOAdvVirgoFixdD1Gpc}, it may help to track the mean errors listed in  
Table~\ref{tab:ParamEstimErrorsNetwork}. Studying these plots and numbers reveals some interesting 
aspects of these signals. First, for the PN waveforms the distance error improves only slightly in 
going from an SNR of 10 to a source distance of 1Gpc. This is easily explained by the fact that the 
sky-averaged SNR of these systems at $d_L$=1Gpc is only slightly greater than 10. Second, the distance error 
reduces a little for complete waveforms \emph{vis \`{a} vis} inspiral ones at 1Gpc. This is mainly due to the 
increased SNR of the former. Third, the error for the complete waveforms for the $M=100M_\odot$ system 
at 1Gpc is still the smallest of all the cases studied here because its sky-averaged SNR is sufficiently large; 
indeed, it is large enough to even compensate for the increased covariance between $\varphi_0$ and 
$(\psi,\iota)$ arising from the merger and ringdown phases, as discussed above.

\begin{table}[tbh]
    \begin{center}
        \begin{tabular}{ccccccccccccccccccc}
            \hline
            \hline
            &\vline& \multicolumn{2}{c}{$\rho = 10$} &\vline& \multicolumn{2}{c}{$d_L=1\mathrm{Gpc}$} \\
            \hline
            $M/M_\odot$ &\vline& \multicolumn{1}{c}{$\Delta \Omega$} & \multicolumn{1}{c}{$\Delta d_L/d_L$} 
            & \vline & \multicolumn{1}{c}{$\Delta \Omega$} & \multicolumn{1}{c}{$\Delta d_L/d_L $}\\
            \hline
            20 & \vline & 0.78\,(2.2) & 55.7\%\,(55.3\%)   &\vline& 0.70\,(2.1) & 43.2\%\,(46.8\%) \\
            100 & \vline & 0.55\,(8.9)  & 111\%\,(63.1\%)   &\vline& 0.13\,(5.9) & 23.0\%\,(39.8\%) \\
            \hline
            \hline
        \end{tabular}
        \caption{Sky-averaged errors in estimating $\Omega$ and $d_L$ using complete BBH 
        waveforms in the case of AdvLIGO-AdvVirgo network. The left column tabulates the errors 
        corresponding to a fixed value $\rho=10$ for the network SNR, while the right column tabulates the
        errors corresponding to a fixed value $d_L=\mathrm{1Gpc}$ of the luminosity distance.
        Errors computed using PN templates are shown in parentheses. The $\Omega$ errors are given in 
        square degrees and the fractional $d_L$ errors are given in percentage.}
        \label{tab:ParamEstimErrorsNetwork}
    \end{center}
\end{table}

Finally, we compare our results with a couple of past studies in the form of
Refs.~\cite{Cutler:1994ys,Jaranowski:1996hs}. First, both these early studies 
used the same noise PSD for both LIGO and Virgo detectors. Second, their
noise PSD was different from both the AdvLIGO and the AdvVirgo noise PSDs used
here; it made their detectors more sensitive (by a factor of a few in amplitude)
in the band below 70Hz and somewhat less sensitive at higher frequencies than
the AdvLIGO PSD used here. Third, they considered only {\em inspiral} signals 
from binary neutron stars with a component mass of 1.4$M_\odot$, and distributed 
them uniformly across a spatial volume. Fourth, in Ref.~\cite{Jaranowski:1996hs}
the authors culled every source that gave a distance error of greater than 100\% or 
that had an SNR of less than 8.5. In our study, where all sources were kept at a fixed 
distance of 1Gpc, none of them were culled. Also, whereas all our sources with 
$M=100~M_\odot$ have an SNR greater than about 25, those with $M=20~M_\odot$ 
have the smallest SNR equal to 6. These differences make it difficult to compare 
these different studies. It is, however, possible to make some limited comparisons. 
Specifically, Fig.~15 in Ref.~\cite{Jaranowski:1996hs} suggests that the 
fractional errors in the estimated source distances all tend to be greater than 
100\% as their source distance approaches 1Gpc. Figure 14 of 
Ref.~\cite{Cutler:1994ys} depicts a similar trend. This appears to be 
consistent with our numbers.

\section{Summary}
\label{sec:Summary}

In this paper, we studied the statistical errors in estimating the parameters
of non-spinning BH binaries using ground-based GW observatories. Our study was 
restricted to the leading harmonic of the GW polarizations of such sources; but 
employing waveforms modelling the inspiral, merger and ring-down stages of the 
binary coalescence. We obtain results both for single- and multi-detector searches.
The single-detector problem was investigated in the context of two generations of 
ground-based detectors, namely, Initial LIGO and Advanced LIGO, as well as
Enhanced LIGO, with intermediate sensitivity. On the other hand,
the multi-detector problem was investigated in the context of the Advanced LIGO-Advanced 
Virgo network. For these calculations, we adopted a two-pronged approach: We first analytically 
computed the error bounds using the Fisher-matrix formalism. We then pointed out the limitations of 
this approach and improved upon those calculations by full-fledged Monte-Carlo simulations.   

To summarize, we find that with an Advanced LIGO detector the total mass of an equal-mass 
binary with $M=20 M_\odot [100 M_\odot]$ located at 1 Gpc can be estimated 
with an accuracy of $\sim 0.67 [\sim 0.34]\%$, while its symmetric mass ratio can be estimated 
with an accuracy of $\sim 1.26 [\sim 0.84]\%$. The effective distance can be estimated with 
an accuracy of $\sim 4.87 [\sim 1.36]\%$ and the time-of-arrival can be placed within $\sim 0.11 [\sim 0.46]$ 
ms. We considered binaries with three different mass ratios ($\eta = 0.25, 0.2222, 0.16$) in 
the range $10 M_\odot \leq M \leq 450 M_\odot$ for these calculations. These results predict 
for a significantly more accurate astrophysical characterization than what has been presented 
in the past literature (which use the 
post-Newtonian waveforms intended to model only the inspiral stage of the binary). To wit, 
the error-bounds for total mass, computed using the complete waveforms is better 
than those computed using the inspiral-only waveforms by a factor of $\sim 1.4\, [\sim 16]$ 
for an equal-mass binary with total mass $20 M_\odot [100 M_\odot]$. The error-bounds on the symmetric
mass ratio is improved by a factor of $\sim 1.4\, [\sim 15]$, those on the time-of-arrival
is improved by a factor of $\sim 7\, [\sim 34]$ and those on the effective distance
is improved by a factor of $\sim 1.1\, [\sim 4]$ by the inclusion of the merger and ringdown stages.

In the case of a network consisting of two Advanced LIGO detectors and one Advanced 
Virgo detector, we found that the luminosity distance to an equal-mass binary with $M=20M_\odot$ 
at 1Gpc can be estimated with a sky- and orientation-averaged accuracy of 43.2\% and the sky 
location can be estimated with a mean accuracy of 0.7 square degrees. For a similar binary, but 
with $M=100M_\odot$, the respective mean accuracies are 23\% and 0.13 square degrees.
For low-mass binaries, with $(M\sim20M_\odot)$, the improvement in the sky-position accuracy due 
to the inclusion of merger and ringdown is about a factor of 3, while for high-mass binaries 
$(M\sim100M_\odot)$, that improvement is by a factor of 45. The inclusion of the same two phases 
betters the distance estimates by a few (for low mass systems) to several (for high mass systems) 
percent. In short, the sky resolution is greatly improved by the inclusion of merger and ringdown, 
while the improvement in the estimation of the luminosity distance arises largely from the extra SNR 
contributed by the merger and ringdown. 

In the case of the AdvLIGO-AdvVirgo detector network, the parameter-estimation 
accuracy peaks for binaries with $M \simeq 100 M_\odot$. Although the observational evidence
for BHs in this mass range is only suggestive, there is growing consensus in the astronomy
community that IMBHs could exist in dense stellar clusters. The existence of 
this class of black holes could explain a number of observations, such as the ultraluminous
X-ray sources and the excess dark matter concentration in globular clusters. 

Several authors have considered the scenario of the coalescence of IMBHs
and have come up with coalescence-rate predictions~\cite{Mandel:2007hi,AmaroSeoane:2006py}. 
Particularly interesting is the case of the merger of two stellar clusters each hosting 
an IMBH considered in Ref.~\cite{AmaroSeoane:2006py}. Since this is expected
to be a strong source of GW signal with a possible EM counterpart~\footnote{It must 
be pointed out that the nature of the EM counterpart is not very clear at the moment.},  
it is a worthwhile question to ask what kind of constraints can be put on the values of cosmological 
parameters by combining GW and EM observations of such sources~\cite{Schutz86}. The improved 
parameter estimation might help to tighten these constraints. This is being investigated in an ongoing
work~\cite{BBH-Cosmology}.  

\acknowledgments 
We would like to thank Giovanni Losurdo and Rana Adhikari for providing the projected noise PSDs 
of Advanced Virgo and Enhanced LIGO, respectively. We also thank B. S. Sathyaprakash and 
Christian R\"over for useful comments on the manuscript, and K.~G.~Arun for helpful discussions. 
SB would like to thank Bruce Allen for his warm hospitality during his stay at Hannover. Computations 
reported in this paper were performed with the aid of the Morgane and Atlas clusters of the Albert 
Einstein Institute. This work is supported in part by the NSF grants PHY-0239735 and
PHY-0758172.

\bibliography{References}

\end{document}

%% file: ParamEstimSingDet.table
\begin{table*}[tbh]
    \begin{center}
        \begin{tabular}{ccccccccccccccccccccccc}
            \hline
            \hline
            &\vline& \multicolumn{3}{c}{$\eta = 0.25$} &\vline& \multicolumn{3}{c}{$\eta = 0.2222$} &\vline& \multicolumn{3}{c}{$\eta = 0.16$} \\
            \hline
            $\rho$ &\vline& \multicolumn{1}{c}{$M = 20 M_\odot$} & \multicolumn{1}{c}{$100 M_\odot$} & \multicolumn{1}{c}{$200 M_\odot$} 
            &\vline& \multicolumn{1}{c}{$20 M_\odot$} & \multicolumn{1}{c}{$100 M_\odot$} & \multicolumn{1}{c}{$200 M_\odot$}
            &\vline& \multicolumn{1}{c}{$20 M_\odot$} & \multicolumn{1}{c}{$100 M_\odot$} & \multicolumn{1}{c}{$200 M_\odot$} \\
            \hline
            \hline
            \multicolumn{13}{c}{$\Delta M/M \times 100$} \\
            \hline
            6   &\vline&	2.30 (1.52) &	4.35 (3.80) &	10.0 (5.59) &\vline&	2.12 (2.15) &	4.37 (4.31) &	8.50 (6.58)  &\vline&	1.66 (1.80) &	4.59 (4.58) &	5.11 (5.74) \\	
            10  &\vline&	1.38 (0.93) &	2.61 (2.39) &	6.02 (3.57)  &\vline&	1.27 (1.50) &	2.62 (2.60) &	5.10 (4.49)  &\vline&	1.00 (1.16) &	2.75 (2.58) &	3.07 (3.61) \\	
            20  &\vline&	0.69 (0.46) &	1.30 (1.11) &	3.01 (1.94)  &\vline&	0.64 (0.76) &	1.31 (1.20) &	2.55 (2.73)  &\vline&	0.50 (0.60) &	1.38 (1.21) &	1.53 (1.81) \\	
            40  &\vline&	0.34 (0.19) &	0.65 (0.51) &	1.50 (0.95)  &\vline&	0.32 (0.34) &	0.66 (0.56) &	1.28 (1.33)  &\vline&	0.25 (0.25) &	0.69 (0.61) &	0.77 (0.83) \\	
            100 &\vline&	0.14 (0.07) &	0.26 (0.18) &	0.60 (0.31)  &\vline&	0.13 (0.11) &	0.26 (0.20) &	0.51 (0.46)  &\vline&	0.10 (0.08) &	0.28 (0.24) &	0.31 (0.28) \\	
            \hline
            \multicolumn{13}{c}{$\Delta \eta/\eta \times 100$} \\
            \hline
            6   &\vline&	4.30 (2.51) &	10.9 (4.40) &	18.8 (6.93) &\vline&	3.94 (3.98) &	11.5 (7.25) &	20.2 (10.2) &\vline&	3.06 (3.29) &	12.5 (9.98)  &	23.6 (15.4) \\	
            10  &\vline&	2.58 (1.77) &	6.54 (3.13)  &	11.3 (4.96) &\vline&	2.36 (2.96) &	6.87 (5.26)  &	12.1 (9.58)  &\vline&	1.84 (2.27) &	7.53 (6.64)   &	14.2 (12.1) \\	
            20  &\vline&	1.29 (0.88) &	3.27 (1.54)  &	5.64 (2.93)  &\vline&	1.18 (1.45) &	3.44 (3.01)  &	6.06 (6.43)   &\vline&	0.92 (1.13) &	3.76 (3.44)   &	7.07 (7.25)   \\	
            40  &\vline&	0.64 (0.36) &	1.63 (0.69)  &	2.82 (1.48)  &\vline&	0.59 (0.65) &	1.72 (1.45)  &	3.03 (3.25)   &\vline&	0.46 (0.47) &	1.88 (1.69)   &	3.54 (3.63)   \\	
            100 &\vline&	0.26 (0.14) &	0.65 (0.29)  &	1.13 (0.52)  &\vline&	0.24 (0.21) &	0.69 (0.53)  &	1.21 (1.16)   &\vline&	0.18 (0.15) &	0.75 (0.67)   &	1.41 (1.33)   \\
            \hline
            \multicolumn{13}{c}{$\Delta \mc/\mc \times 100$} \\
            \hline
            6   &\vline&	0.32 (0.94) &	4.56 (3.86) &	19.8 (7.92) &\vline&	0.28 (1.16) &	4.34 (4.64) &	19.0 (10.6) &\vline&	0.20 (1.16) &	3.85 (4.22) &	16.5 (12.0) \\	
            10  &\vline&	0.19 (0.31) &	2.73 (2.62) &	11.9 (5.56) &\vline&	0.17 (0.35) &	2.60 (3.20) &	11.4 (9.30)  &\vline&	0.12 (0.26) &	2.31 (2.76) &	9.92 (9.51) \\
            20  &\vline&	0.10 (0.10) &	1.37 (1.24) &	5.95 (3.23)  &\vline&	0.08 (0.13) &	1.30 (1.65) &	5.70 (6.24)   &\vline&	0.06 (0.09) &	1.15 (1.45) &	4.96 (5.52) \\	
            40  &\vline&	0.05 (0.03) &	0.68 (0.55) &	2.97 (1.64)  &\vline&	0.04 (0.05) &	0.65 (0.68) &	2.85 (3.14)   &\vline&	0.03 (0.03) &	0.58 (0.59) &	2.48 (2.67) \\	
            100 &\vline&	0.02 (0.01) &	0.27 (0.20) &	1.19 (0.57)  &\vline&	0.02 (0.01) &	0.26 (0.26) &	1.14 (1.11)   &\vline&	0.01 (0.01) &	0.23 (0.23) &	0.99 (0.97) \\	
            \hline
            \hline
            \multicolumn{13}{c}{$\Delta d_{\rm eff}/d_{\rm eff} \times 100$} \\
            \hline
            6   &\vline&	16.7 &	17.6 &	24.6 &\vline&	16.7 &	17.5 &	24.3 &\vline&	16.7 &	17.4 &	23.2 \\	
            10  &\vline&	10.0 &	10.5 &	14.8 &\vline&	10.0 &	10.5 &	14.6 &\vline&	10.0 &	10.4 &	13.9 \\	
            20  &\vline&	5.00  &	5.27  &	7.39 &\vline&	5.00  &	5.26 &	7.30 &\vline&	5.00  &	5.21  &	6.95 \\	
            40  &\vline&	2.50  &	2.63  &	3.70 &\vline&	2.50  &	2.63 &	3.65 &\vline&	2.50  &	2.61  &	3.48 \\	
            100 &\vline&	1.00  &	1.05  &	1.48 &\vline&	1.00  &	1.05 &	1.46 &\vline&	1.00  &	1.04  &	1.39 \\	
            \hline
            \multicolumn{13}{c}{$\Delta t_0 ({\rm ms})$} \\
            \hline
            6   &\vline&	0.37 &	5.90 &	15.8 &\vline&	0.39 &	7.22  &	20.0 &\vline&	0.42 &	11.7 &	36.3 \\	
            10  &\vline&	0.22 &	3.54 &	9.47  &\vline&	0.23 &	4.33  &	12.0 &\vline&	0.25 &	7.04  &	21.8 \\	
            20  &\vline&	0.11 &	1.77 &	4.73  &\vline&	0.12 &	2.17  &	6.00  &\vline&	0.13 &	3.52  &	10.9 \\	
            40  &\vline&	0.06 &	0.88 &	2.37  &\vline&	0.06 &	1.08  &	3.00  &\vline&	0.06 &	1.76  &	5.45 \\	
            100 &\vline&	0.02 &	0.35 &	0.95  &\vline&	0.02 &	0.43  &	1.20  &\vline&	0.03 &	0.70  &	2.18 \\	
            \hline
            \hline
        \end{tabular}
        \caption{Errors in estimating different prameters of the binary, as a function of the binary parameters
        and the SNR, computed using the Fisher matrix formalism. Same errors computed from the Monte-Carlo
        simulations are shown in brackets. }
        \label{tab:ParamEstimErrorsAdvLIGO}
    \end{center}
\end{table*}